\documentclass[a4paper,fleqn,usenatbib]{mnras}

\usepackage{newtxtext,newtxmath}
\usepackage[T1]{fontenc}
\usepackage{ae,aecompl}

\usepackage{graphicx}
\usepackage{amsmath}
\usepackage{amssymb}
\usepackage{bm}
\usepackage{comment}
\hypersetup{draft}

\newcommand{\beq}{\begin{equation}}
\newcommand{\eeq}{\end{equation}}
\newcommand{\beqa}{\begin{eqnarray}}
\newcommand{\eeqa}{\end{eqnarray}}
\newcommand{\vir}{\mathrm{vir}}
\newcommand{\Mpc}{\mathrm{Mpc}}

\newcommand{\Msun}{\mathrm{M}_\odot}
\newcommand{\rmhm}{\mathrm{hm}}

\title[Orientation dependence of mass profiles]
{Strong orientation dependence of surface mass density profiles of
dark haloes at large scales}

\author[K. Osato et al.]{
Ken Osato,$^{1}$\thanks{E-mail: ken.osato@utap.phys.s.u-tokyo.ac.jp}
Takahiro Nishimichi,$^{2}$
Masamune Oguri,$^{1,2,3}$
Masahiro Takada,$^{2}$
\newauthor
and Teppei Okumura$^{4,2}$
\\
$^{1}$Department of Physics, School of Science, The University of Tokyo,
7-3-1 Hongo, Bunkyo, Tokyo, 113-0033, Japan\\
$^{2}$Kavli Institute for the Physics and Mathematics of the Universe (WPI),
U-Tokyo Institutes for Advanced Study,\\
The University of Tokyo, Kashiwa, Chiba, 277-8583, Japan\\
$^{3}$Research Center for the Early Universe, The University of Tokyo,
7-3-1 Hongo, Bunkyo, Tokyo, 113-0033, Japan\\
$^{4}$Institute of Astronomy and Astrophysics, Academia Sinica, Taipei 10617, Taiwan
}

\date{Accepted XXX. Received YYY; in original form ZZZ}

\pubyear{2017}

\begin{document}
\label{firstpage}
\pagerange{\pageref{firstpage}--\pageref{lastpage}}
\maketitle

\begin{abstract}
We study the dependence of surface mass density profiles, which can be
directly measured by weak gravitational lensing, on the orientation of
haloes with respect to the line-of-sight direction, using
a suite of $N$-body simulations. We find that, when major axes of haloes are
aligned with the line-of-sight direction, surface mass density
profiles have higher amplitudes than those averaged over all halo
orientations, over all scales from $0.1$ to $100\,\Mpc/h$ we studied.
While the orientation dependence at small scales is ascribed to
the halo triaxiality, our results indicate even stronger
orientation dependence in the so-called two-halo regime, up to $100\,\Mpc/h$.
The orientation dependence for the two-halo term is
well approximated by a multiplicative shift of the amplitude
and therefore a shift in the halo bias parameter value.
The halo bias from the two-halo term can be overestimated or underestimated
by up to $\sim 30\%$ depending on the
viewing angle, which translates into the bias in estimated halo masses
by up to a factor of two from halo bias measurements.
The orientation dependence at large scales originates
from the anisotropic halo-matter correlation function,
which has an elliptical shape with the axis
ratio of $\sim 0.55$ up to $100\,\Mpc/h$.
We discuss potential impacts of halo orientation bias
on other observables such as optically selected
cluster samples and a clustering analysis of large-scale structure
tracers such as quasars.
\end{abstract}




\section{Introduction}
\label{sec:introduction}
Weak gravitational lensing provides an important means of measuring
matter distributions around galaxies and clusters
\citep[e.g.,][for a review]{Bartelmann2001}.
In particular, cross-correlating galaxy shapes with positions of foreground
galaxies or clusters, which is referred to as stacked weak lensing or galaxy-galaxy
lensing, allows one to measure the average matter distribution
around the foreground tracers, which can be in turn used to
constrain the relation between luminous and matter distributions
in the large-scale structure
\citep[e.g.,][]{Brainerd1996,Sheldon2004A,Mandelbaum2006,Okabeetal:10,
Leauthaud2012,Okabe2013,Umetsu2014,Miyatake2015,Murata2018,Prat2017}.
Combining the stacked lensing with the auto-correlation function of the same
foreground tracers can be used to constrain cosmological parameters
to calibrate the bias uncertainty simultaneously
\citep[e.g.,][]{Baldauf2010,OguriTakada:11,More2015}.

The stacked lensing profile of galaxies or clusters
consists of two distinct regimes, the {\it one-halo} and
{\it two-halo} terms in the halo model picture \citep[e.g.,][for a review]{Cooray2002}.
Here the one-halo term arises from matter distribution
in the same halo, and the lensing signal can be used to constrain
the interior mass, i.e. the total
mass of gravitationally-bound matter in the haloes.
On the other hand, the two-halo term arises from matter in different haloes, more generally
matter distributed in the large-scale structure. The two-halo term contains
cleaner information on the matter distribution that is relatively easier to theoretically
model and is less affected by baryonic physics, which can be used to infer the halo bias
including the assembly bias
\citep[e.g.,][]{Covone2014,Lin2016,Miyatake2016,More2016,Dvornik2017,Busch2017}
and/or the underlying matter power spectrum
\citep[e.g.,][]{Lombriser2012,Mandelbaum2013,Sereno2015,More2015,vanUitert2017b,DES2017}.
If the foreground tracers are
selected in an unbiased manner such that the matter distribution
around the tracers is statistically isotropic,
the stacked weak lensing profile is computed from a
projection of the spherically averaged three-dimensional
halo-matter cross-correlation function at the redshift of foreground tracers
\citep{Mandelbaum2005,Hikageetal:13,More2015}.

However, if the sample of foreground tracers has selection effects
depending on the line-of-sight direction,
referred to as ``projection effects'' in this paper,
it violates the spherical symmetry and could cause a bias in parameters
estimated from the weak lensing measurements.
For instance, it is often advocated that
strong lensing clusters (clusters found via observations of strong lensing phenomena)
are affected by selection effects depending
on an alignment of the major axis of mass distribution to the line-of-sight direction
\citep{Hennawi2007,Oguri2009,Meneghetti2010},
because projected surface mass density profiles are sensitive to the
orientation of dark haloes
\citep{Clowe2004,Oguri2005,Gavazzi2005,Corless2007,Limousin2013}.
This originates from the fact that dark
matter haloes are highly non-spherical,
but rather triaxial with a typical major-to-minor axis ratio
of $\sim 0.5$ for cluster-scale haloes, reflecting the collision-less
nature of dark matter
\citep{Jing2002,Allgood2006,Schneider2012,Vera-Ciro2014,Vega-Ferrero2017},
which has also been confirmed directly by weak lensing observations
\citep{Evans2009,Oguri2010,Oguri2012,Clampitt2016,vanUitert2017a,Shin2018}.
Hence, the line-of-sight projection of the matter distribution for such a sample
under the selection effects complicates an interpretation of the
lensing observables compared to the case of spherical symmetry.
These effects need to be properly taken into account in order to derive
an unbiased estimation of the model parameters as well as to use the
weak lensing observables to constrain cosmological parameters.

In contrast, it is not well understood how projection effects change
surface mass density profiles of dark haloes, especially at large scales
in the two-halo term regime.
It is known that three-dimensional mass distributions
around dark haloes correlate with their shapes,
the so-called intrinsic alignment
\citep[e.g.,][for a review]{Troxel2015,Joachimi2015}.
This intrinsic alignment implies that there is a larger amount of matter
along the major axis of a halo and a smaller amount of matter along the
minor axis of a halo than the spherical average,
and the correlation extends out to very large
scales. However, it is not obvious how the intrinsic alignment affects
{\it projected} mass distributions, i.e., surface mass density
profiles, as a function of the viewing angle of haloes, because the
larger amount of matter along the major axis would be compensated by
the smaller amount of matter along the minor axis when the matter
distribution is projected along the line-of-sight.

In this paper, we study the dependence of surface mass density
profiles around dark haloes on viewing angle, focusing on
cluster-scale dark haloes. Specifically, we use high resolution
$N$-body simulations to quantify the orientation effect on surface
mass density profiles from small to large scales up to
$100\,\Mpc/h$. We discuss its impact on measurements of halo biases
and halo masses from large scale surface mass density profile
measurements. Our results will have an important implication for the
analysis of ongoing surveys such as Dark Energy Survey \citep{DES2016}
and Hyper Suprime-Cam survey \citep{Aihara2018}.

This paper is organized as follows.
We first review the halo model calculation in Section~\ref{sec:formalism}.
In Section~\ref{sec:simulations}, the details of
$N$-body simulations and numerical methods are explained.
We present the measurements of surface mass density profiles
and the fitting results in Section~\ref{sec:results}, and
discuss the origin of the orientation effect in Section~\ref{sec:origin}.
In Section~\ref{sec:discussions}, we discuss implications of our
results from $N$-body simulations on the analysis of actual
observations. Finally, we give our conclusion in Section~\ref{sec:conclusion}.
Throughout this paper, we adopt the flat $\Lambda$ cold dark matter
model with $H_0 = 67.27\,\mathrm{km}/\mathrm{s}/\Mpc$,
$\Omega_\mathrm{m} = 0.3156$, $\Omega_\mathrm{b} = 0.04917$,
$n_\mathrm{s} = 0.9645$, and $A_\mathrm{s} = 2.2065 \times 10^{-9}$
at a pivot wavenumber $k_\mathrm{piv} = 0.05 \, \Mpc^{-1}$
based on {\it Planck} measurements of the anisotropy of
temperature and polarization (TT, TE, EE+lowP) of cosmic microwave
background \citep{Planck2016}.

\section{Halo model}
\label{sec:formalism}
In this Section, we will briefly review the halo model calculation of
surface mass density profiles around haloes, which will be used to
estimate possible biases in estimating halo masses from
the surface mass density profiles simulated from $N$-body simulations.

\subsection{Halo density profile}
The density profile characterizes the structure of dark haloes.
Using $N$-body simulations, \citet{Navarro1996,Navarro1997} proposed a
universal form of the spherically averaged radial density profile, the
so-called Navarro--Frenk--White (NFW) profile
\beq
\rho_\mathrm{h}(r) = \frac{\rho_s}{(r/r_s)(1+r/r_s)^2},
\eeq
where $\rho_s$ is the scale density and $r_s$ is the scale radius.
The virial mass is defined as
\beq
M_\vir = \frac{4\pi}{3} \Delta_\vir \bar{\rho}_\mathrm{cr}(z) r_\vir^3,
\eeq
where $\Delta_\vir$ is the virial overdensity,
$\bar{\rho}_\mathrm{cr}(z)$ is the critical density, and
$r_\vir$ is the virial radius.
For $\Delta_\vir$, we use the following formula based on spherical
collapse model \citep{Bryan1998}
\beq
\Delta_\vir = 18 \pi^2 + 82 \left[\Omega_\mathrm{m} (z)-1\right] -
39 \left[\Omega_\mathrm{m} (z)-1\right]^2 ,
\eeq
where $\Omega_\mathrm{m} (z)$ is matter density at redshift $z$
\beq
\Omega_\mathrm{m} (z) = \Omega_\mathrm{m} (1+z)^3
[\Omega_\mathrm{m} (1+z)^3 + \Omega_\Lambda]^{-1} .
\eeq
The scale density can be determined from the virial mass as
\beq
M_\vir = \int_0^{r_\vir} \!\! \rho_\mathrm{h}(r) 4\pi r^2 \mathrm{d}r =
4 \pi \rho_s r_s^3 m_\mathrm{NFW} (c_\vir) ,
\eeq
where
\beq
m_\mathrm{NFW} (c_\vir) = \int_0^{c_\vir} \!\!
\frac{x}{(1+x)^2} \mathrm{d}x = \ln (1+c_\vir) -\frac{c_\vir}{1+c_\vir}.
\eeq
The parameter $c_\vir$ is the concentration parameter, which is the ratio of
the virial radius to the scale radius, i.e., $c_\vir = r_\vir/r_s$.
Previous studies \citep[e.g.,][]{Duffy2008,Diemer2015} based on
$N$-body simulations have shown that
the mean relation of the concentration parameter is tightly correlated
with halo mass and redshift,
and there is a sizable scatter around the mean relation,
typically by an amount of $\sigma_{\ln c} \sim 0.2$ for cluster-scale haloes.

The NFW profile has an asymptotic form as $\rho_\mathrm{h} \propto r^{-3}$ for
$r \gg r_s$, which indicates that the total enclosed halo mass diverges
at $r \rightarrow \infty$.
Thus, in the halo model calculations the prescription to truncate the
NFW density profile at the virial radius is commonly used
\citep{Takada2003a,Takada2003b}.
However, this truncation produces an unphysical discontinuity
in the weak lensing profiles that are obtained from the line-of-sight projection
of an isolated NFW profile.
In order to avoid this problem, \citet{Baltz2009} proposed
another density profile (hereafter the BMO profile) which has the
steeper asymptotic form at outer radii
\beq
\rho_\mathrm{h}(r) = \frac{\rho_s}{(r/r_s)(1+r/r_s)^2}
\left( \frac{r_t^2}{r^2+r_t^2} \right)^n ,
\eeq
where $r_t$ is the truncation radius. Based on the comparison of halo
model calculations with ray-tracing simulations,
\citet{Oguri2011} proposed to adopt the following value for the
truncation radius for $n=2$;
\beq
r_t = 2.7 r_\vir,
\eeq
which we also adopt in this paper.
The BMO profile with $n=2$ has a steep outskirts ($\propto r^{-7}$),
and as a result the total enclosed mass converges quickly.
Throughout this paper, we use the BMO profile as a fiducial density
profile of dark haloes.

\subsection{Surface mass density profile}
Under the assumption that all matter in the Universe is associated with haloes,
we can analytically compute the surface mass density profile from
small to large scales using the halo model \citep{Guzik2002,Mandelbaum2005}.
In this framework, the surface mass density profile as a function of
projected radius $R$, $\Sigma(R)$, can be decomposed into the one-halo
term $\Sigma^\mathrm{1h} (R)$ and the two-halo term $\Sigma^\mathrm{2h} (R)$ as
\beqa
\Sigma (R) &=& \Sigma^\mathrm{1h} (R) + \Sigma^\mathrm{2h} (R),
\label{eq:halomodel}\\
\Sigma^\mathrm{1h} (R) &=& \int_{-\infty}^\infty \!\! \mathrm{d}\Pi \,
\rho_\mathrm{h}\left( \sqrt{R^2+\Pi^2} \right) ,
\label{eq:sig1h}\\
\Sigma^\mathrm{2h} (R) &=&
\bar{\rho}_\mathrm{m0}b
\int_0^\infty \!\! \frac{k\mathrm{d}k}{2\pi} \, P_\mathrm{lin}(k ; z) J_0(kR) ,
\label{eq:sig2h}
\eeqa
where $\bar{\rho}_\mathrm{m0}$ is the mean matter density in the present Universe,
$b$ is the linear halo bias, $J_0 (x)$ is the zeroth order Bessel function, and
$P_\mathrm{lin}(k ; z)$ is the linear matter power spectrum.
We compute the linear matter power spectrum using the {\tt CAMB} code
\citep{Lewis2000}.
Also see \cite{Takada2003a} and \cite{OguriTakada:11}
for the halo model formulation of lensing profiles
based on an use of the two- and three-dimensional
Fourier transforms of the halo density profile.

In weak lensing observations, we usually measure tangential shear
profiles with respect to the center of each foreground halo,
which measure {\it excess} surface mass density
profiles $\Delta\Sigma(R)$ rather than $\Sigma(R)$.
The excess surface mass density profile is
related to the surface mass density profile as
\beq
\Delta \Sigma(R) = \bar{\Sigma} (<R) - \Sigma (R),
\label{eq:dsigma}
\eeq
where $\bar{\Sigma}(<R)$ is the mean surface density within
a circular aperture of radius $R$.
Since $\Sigma(R)$ and $\Delta\Sigma(R)$ carry the same information,
in this paper we focus only on $\Sigma(R)$.
We note that our results on $\Sigma(R)$ can easily be converted
to those on $\Delta\Sigma(R)$ via equation~\eqref{eq:dsigma}.
Throughout this paper, the surfass mass desity $\Sigma (R)$ denotes
the projected mass density from which the mean density has already been subtracted.

\section{Simulations}
\label{sec:simulations}

\subsection{$N$-body simulations}
In this Section, we describe the details of simulations used in the analysis.
In order to obtain the distributions of matter and haloes in the Universe,
we run $N$-body simulations.
For this purpose, we use Tree-PM code {\tt Gadget-2} \citep{Springel2005}.
The number of particles is $2048^3$ and the length of the simulation box
on a side is $1\,\mathrm{Gpc}/h$. The corresponding particle mass
is $m_\mathrm{particle} = 1.02 \times 10^{10} \Msun/h$.
The initial conditions at the redshift $z_\mathrm{ini} = 60$ are generated with
the parallel code developed in \citet{Nishimichi2009,Nishimichi2010}
and \citet{Valageas2011},
which employs the second-order Lagrangian perturbation theory.
We generated 24 initial conditions with different random seeds
and ran the $N$-body simulations to obtain the matter distribution
at the present Universe, i.e., $z=0$.

\subsection{Halo identification and halo shape measurement}

Next, we run the halo finding algorithm,
the {\tt Rockstar} \citep{Behroozi2013}, in each $N$-body simulation output.
The minimum halo mass which we use in the
analysis is $M_\vir = 10^{14} \Msun/h$. Hence, haloes used in the
analysis contain at least $\sim 10^4$ particles, which is sufficient
for reliable shape measurements of individual haloes \citep{Jing2002}.
Hereafter, the halo samples are divided into three bins,
$M_\vir/[\Msun /h] \in [10^{14}, 5 \times 10^{14}]$,
$[5 \times 10^{14}, 10^{15}]$, and $[10^{15}, 10^{16}]$,
according to their virial masses, respectively.

We then compute inertia tensor for each halo. While there are various
definitions of the inertia tensor \citep[][and references therein]{Bett2012},
in this paper we employ the reduced inertia tensor determined by an iterative scheme
to characterize the shape of dark haloes.
The reduced tensor for a halo with $N$ particles is defined as
\beq
\mathcal{M}_{ij}^{(k)} = \sum_{p=1}^N m_\mathrm{particle}
\frac{R^{(k)}_{p,i} R^{(k)}_{p,j}}{\left( \tilde{R}^{(k)}_p \right)^2},
\label{eq:m_ij}
\eeq
where $\bm{R}^{(k)}_p$ ($\tilde{R}^{(k)}_p$) is
the triaxial coordinate (radius) of the $p$-th member particle
in the halo at the $k$-th step. At the first step ($k = 1$), the triaxial
radius is the same as the one in Cartesian coordinates. At the $k$-th
($k \geq 2$) step, the triaxial radius is measured in the principal
coordinates at the last ($(k-1)$-th) step, i.e.,
\beq
\tilde{R}^{(k)}_p = \left( \frac{X^{(k)}_p}{q^{(k-1)}} \right)^2+
\left( \frac{Y^{(k)}_p}{s^{(k-1)}} \right)^2 + \left( Z^{(k)}_p \right)^2 ,
\eeq
where $X^{(k)}_p$, $Y^{(k)}_p$ and $Z^{(k)}_p$ are coordinates along the minor,
intermediate, and major axis directions, respectively,
$q^{(k)}$ ($s^{(k)}$) is the minor-to-major (intermediate-to-major) axis ratio, and
\beq
\bm{R}^{(k)}_p = \left( X^{(k)}_p, Y^{(k)}_p, Z^{(k)}_p \right) .
\eeq
Since the Cartesian coordinate system is adopted in the first step,
the axis ratios are unity, i.e.,
\beq
q^{(0)} = 1, s^{(0)} = 1 .
\eeq
We iteratively compute the tensor until the fractional difference of
axial ratios between two steps becomes less than $1\%$.
The sum in equation~\eqref{eq:m_ij} runs over all the
particles whose triaxial radii are less than
$r_\vir$. The eigenvalue (eigenvector) of the inertia tensor corresponds to
the length (direction) of the principal axis.

Hereafter, we focus on the direction of the major axis with respect to
the line-of-sight direction, which is chosen to be the
$z$-axis in the simulation box. For each mass bin, we divide the halo
sample based on the directional cosine between the major axis and the
line-of-sight direction, which we denote as $\cos i$.
We divide the halo sample into five bins that are equally spaced with
respect to $\cos i$\footnote{The area element of the unit sphere in
polar coordinates is $\mathrm{d} (\cos \theta) \mathrm{d} \phi$.
Therefore, the number of haloes which belong to each bin is almost
the same when the direction of the major axis is random.}.

\section{Results}
\label{sec:results}
\subsection{Halo-matter cross-spectrum}
\begin{figure}
\includegraphics[width=8cm]{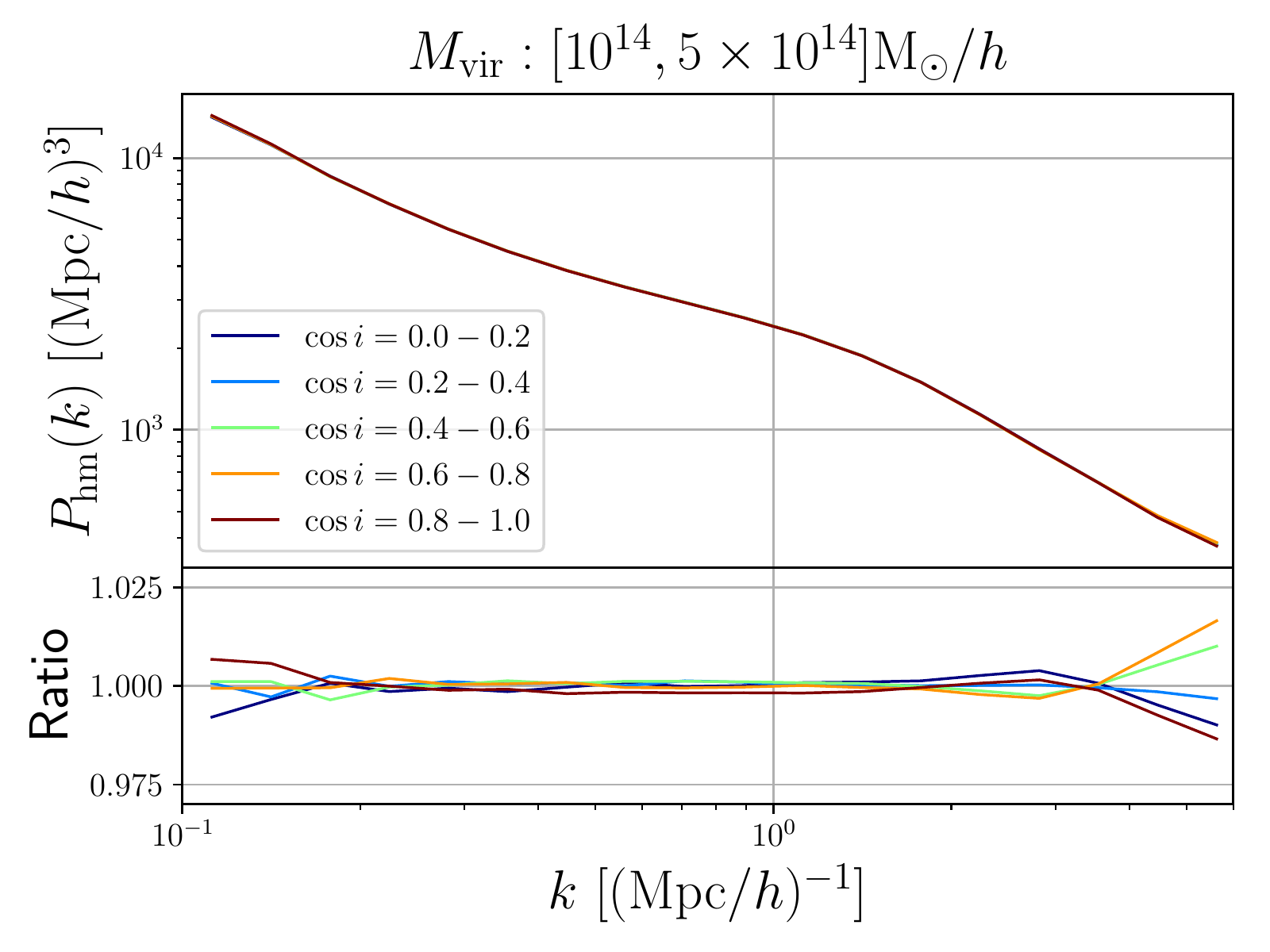}
\includegraphics[width=8cm]{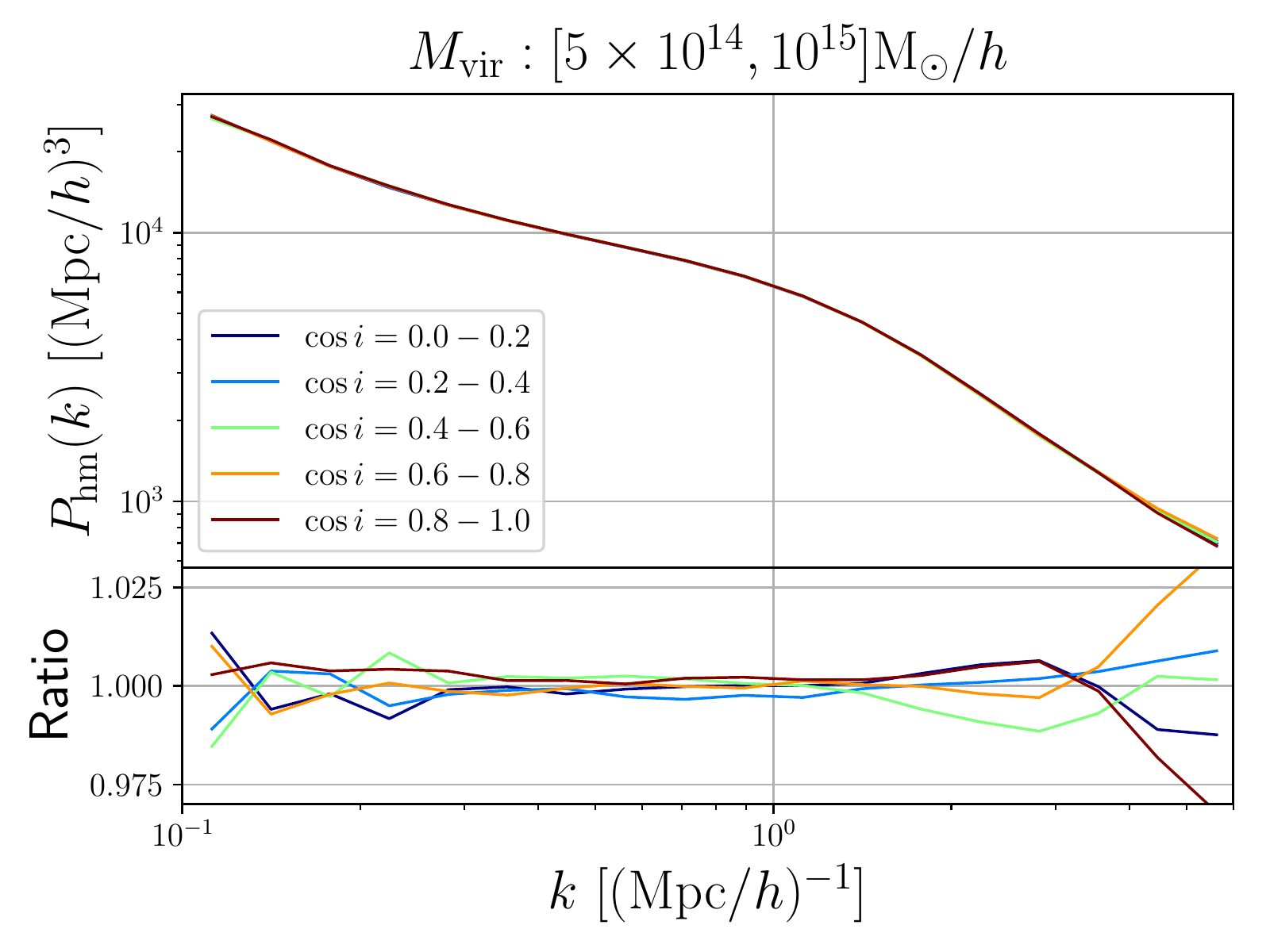}
\includegraphics[width=8cm]{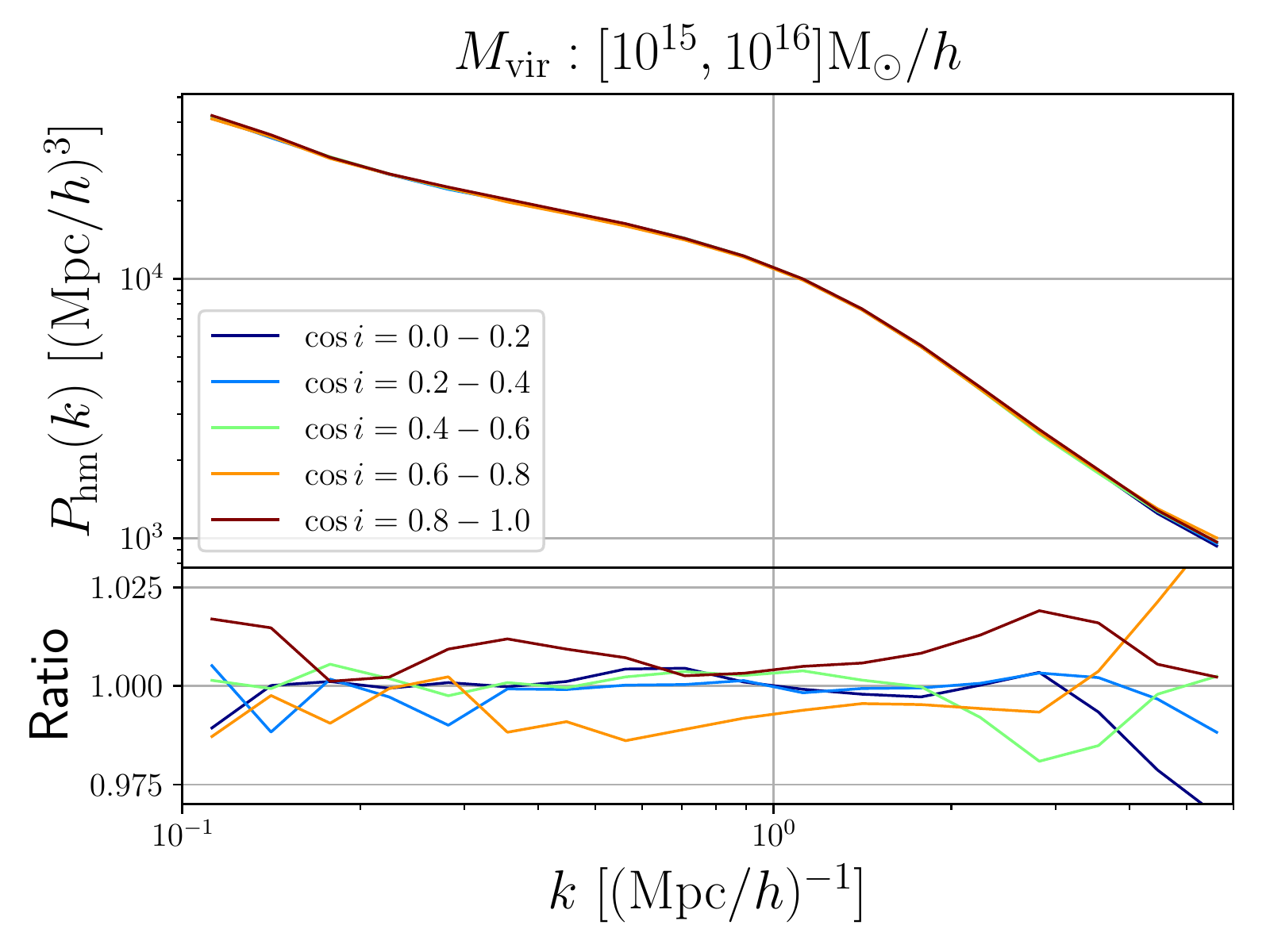}
\caption{The three-dimensional halo-matter cross power spectra measured in
  the simulations. Different lines indicate cross power spectra for
  different halo samples that are binned according to their
  orientation with respect to the line-of-sight direction. The upper,
  middle, and lower plots show the cross power spectra for halo samples with
  mass ranges of $[10^{14}, 5 \times 10^{14}]$, $[5 \times 10^{14}, 10^{15}]$,
  and $[10^{15}, 10^{16}] \Msun /h$, respectively.
  The lower panels show the ratios of cross power spectra for
  individual orientation bins to the cross power spectrum computed
  using all haloes. Since the Universe has no preferred direction, the
  cross power spectra do not depend on the halo orientation, which is
  explicitly confirmed in these plots.}
\label{fig:Phm}
\end{figure}

Since the Universe has no preferred direction, spherically averaged
mass density profiles around dark haloes should not depend on their
orientations. As a sanity check, we calculate three-dimensional
halo-matter cross-spectra in the simulations for the five orientation
bins to make sure that these cross power spectra, which correspond to
spherically averaged mass profiles in real space, do not depend on
the halo orientation. Our results in Figure~\ref{fig:Phm} indeed
confirm that the cross-spectra are consistent with each other among
different orientation bins and the variation is well
below statistical scatters over 24 realizations.

\subsection{Surface mass density profile}
\begin{figure}
\includegraphics[width=8cm]{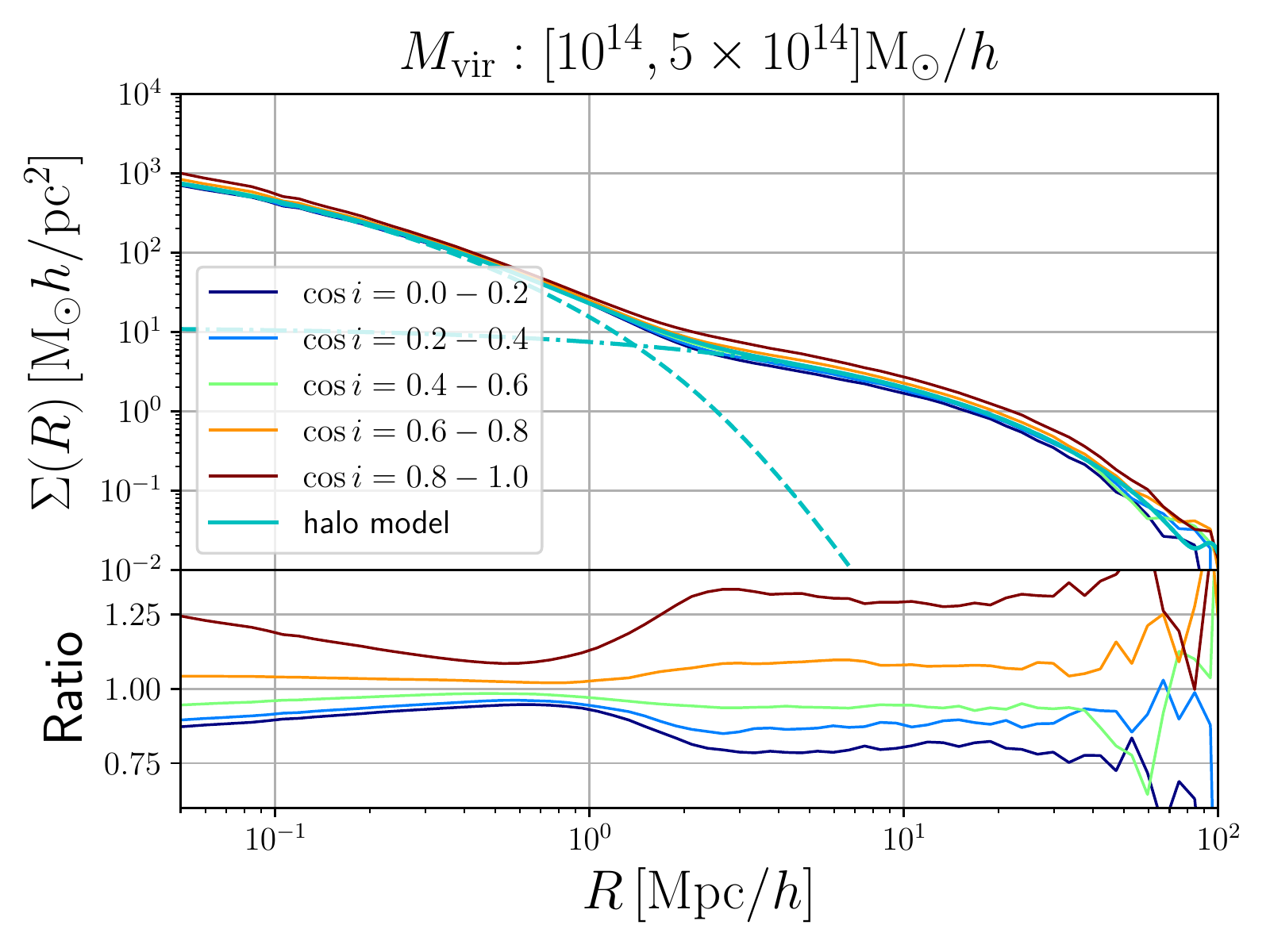}
\includegraphics[width=8cm]{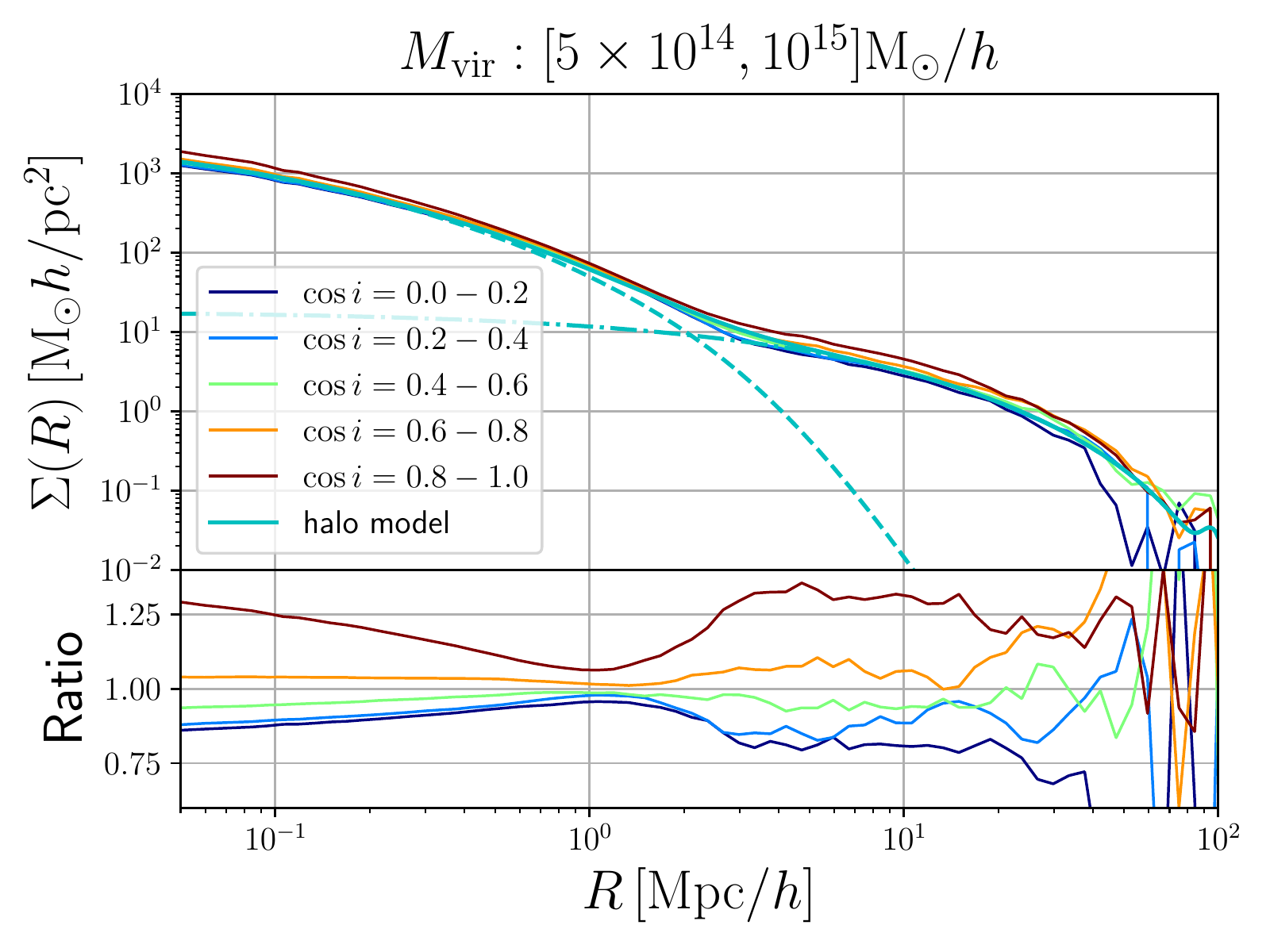}
\includegraphics[width=8cm]{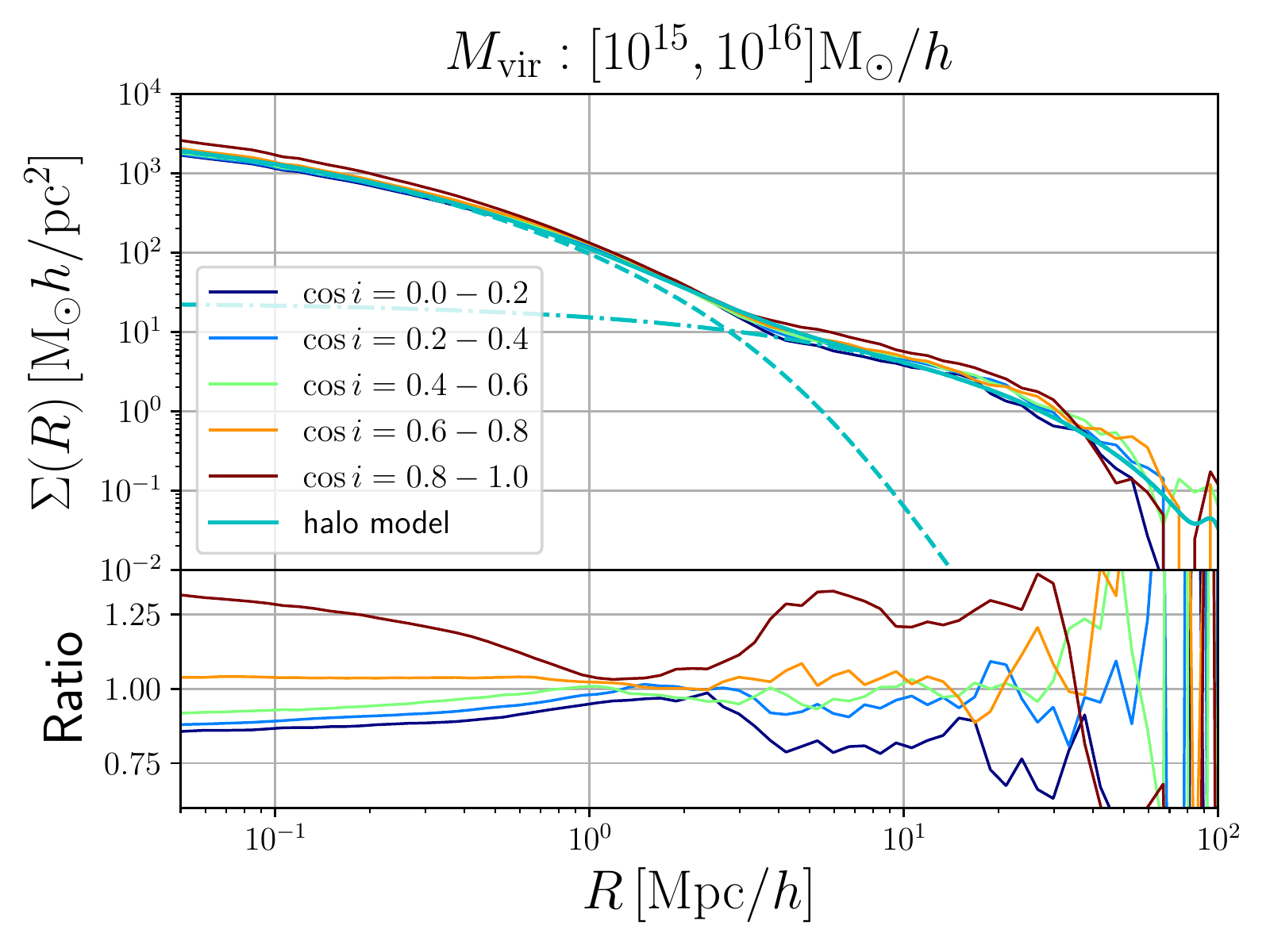}
\caption{Surface mass density profiles measured in the simulations.
  Different solid lines indicate surface mass densities for
  different halo samples that are binned according to their
  orientation with respect to the line-of-sight direction,
  which is parametrized by $\cos i$, the cosine angle between the major axis of
  each halo and the line-of-sight direction. The curves with
  $\cos i\sim 1$ denote the results for halos whose major axes
  are almost perfectly aligned with the line-of-sight.
  The upper, middle, and lower plots show surface mass density
  profiles for halo samples with mass ranges of
  $[10^{14}, 5 \times 10^{14}]$, $[5 \times 10^{14}, 10^{15}]$,
  and $[10^{15}, 10^{16}] \Msun /h$, respectively.
  The lower panels show the ratios of surface mass density profiles for
  individual orientation bins to the surface mass density profile
  averaged over all orientations. The cyan lines show the halo model
  calculation with the best-fit parameters in the case where the
  average surface mass density profiles for all haloes are used.
  The dashed (dot-dashed) line shows the contribution from the
  one-halo (two-halo) term to the best-fit halo model.}
\label{fig:Sigma}
\end{figure}

In Figure~\ref{fig:Sigma}, we show surface mass density profiles for
different orientation bins, as well as those averaged over all halo
orientations. For all mass bins, clear dependence on the orientation
can be seen for a wide range of radii, from $0.1\,\Mpc/h$ up to
$100\,\Mpc/h$. When we observe haloes whose major axes are aligned
with the line-of-sight direction, the amplitudes of the surface mass
density profiles are always larger than the average surface mass
density profile.
The cross-correlation between the intra-halo structure
at a few $\la \Mpc/h$ and the large-scale matter
distribution might be somewhat surprising.
The existence of strong cross-correlation implies
that the matter distributions at the totally different scales
co-evolve in nonlinear structure formation. We will later discuss the nature of
the cross-correlation in more detail.

To study the orientation bias
in the surface mass density profiles more quantitatively,
we fit the surface mass density profiles using the halo model described
in Section~\ref{sec:formalism}. We fit surface mass density profiles
in simulations with equation~\eqref{eq:halomodel}, leaving
$M_\vir$, $c_\vir$, and $b$ as free parameters. For
an estimation of the model parameters,
we use the standard chi-square fitting:
\beq
\chi^2 = \sum_i \frac{[ \Sigma^\mathrm{sim} (R_i;\cos i) -
\Sigma^\mathrm{model} (R_i;M_\vir, c_\vir, b) ]^2}{\sigma^2_\Sigma (R_i)} ,
\label{eq:chi_square}
\eeq
where the fitting range is $0.05 < R/[\Mpc/h] < 30.0$ with
54 equally log-spaced bins, $\Sigma^\mathrm{sim} (R_i;\cos i)$ is the surface
mass density measured in the simulations,
$\Sigma^\mathrm{model} (R_i;M_\vir, c_\vir, b)$ is the halo model
prediction given parameters $(M_\vir, c_\vir, b)$ and
$\sigma^2_\Sigma (R_i)$ is the variance of the surface mass density
over 24 simulations.
In equation~\eqref{eq:chi_square}, we ignore covariance,
i.e. correlation between different bins,
because the number of realization is too small to accurately estimate the covariance.
In the following we study how the best-fit parameters vary with different inputs of the simulated
surface mass density profiles ($\Sigma^{\mathrm{sim}}$) in equation~(\ref{eq:chi_square}),
where we used the same variance at each radial bin.
Hence we believe that our approximation (neglecting
the off-diagonal covariance components) does not largely affect the comparison study.

The best-fit halo model results shown in Figure~\ref{fig:Sigma}
indicate that the radius where the fractional differences between
haloes with different orientations become minimum ($\sim 1\,\Mpc/h$)
roughly corresponds to the transition between the one-halo and two-halo
terms. This implies that the orientation dependences at small and
large radii have different origin.

In fact it has been known that the halo triaxiality is the main cause
of the orientation dependence of the surface mass density profile in
the one-halo regime
\citep{Oguri2005,Gavazzi2005,Corless2007,Limousin2013}.
We explicitly check this by computing surface mass density profiles
for different halo orientations expected by the triaxial halo model of
\citet{Jing2002}. The triaxial density profile is parametrized as
\beq
\rho_{\mathrm{h}}(R) = \frac{\Delta_\mathrm{ce} \bar{\rho}_\mathrm{cr} (z)}{(R/R_0) (1+R/R_0)^2},
\label{eq:halo_triaxial}
\eeq
where $\Delta_\mathrm{ce}$ is the characteristic overdensity that is
related to the virial overdensity $\Delta_\vir$
\citep[see][]{Jing2002}, $R_0$ is the scale radius, and $R$ is the
triaxial radius. The triaxial radius is different from the ordinary
radius in terms of reference coordinate system and is defined as
\beq
R^2 = c^2 \left(\frac{X^2}{a^2} + \frac{Y^2}{b^2} + \frac{Z^2}{c^2}\right),
\eeq
where $X$, $Y$ and $Z$ are Cartesian coordinates in the principal coordinate system
of the triaxial halo, and $a$, $b$, and $c$ is the length of minor, intermediate,
and major axis (i.e., $a \leq b \leq c$), respectively.
We derive surface mass density profiles
of the triaxial halo model following \citet{Oguri2003} and
\citet{Oguri2009}. In particular, for each mass bin we compute surface
mass density profiles as a function of the halo orientation using the
Monte-Carlo method developed in \citet{Oguri2009}. In this method, in
each mass bin, we randomly generate haloes following the halo mass
function of \citet{Bhattacharya2011}, and assign their axis ratios and
concentration parameters following the probability distribution
functions derived in \citet{Jing2002}. We also assign
orientations of these haloes with respect to the line-of-sight
direction assuming they are randomly oriented. We then connect
project triaxial density profiles with corresponding two-dimensional
mass ($M_\mathrm{2D}$) and concentration parameters ($c_\mathrm{2D}$), and use
the $\Sigma^\mathrm{1h}(R)$ for the spherical NFW halo
(equation~\ref{eq:sig1h}) to predict surface mass density profiles of
individual triaxial haloes generated by the Monte-Carlo method.
This result is used to compute surface mass density profiles for each
$\cos i$ bin for the triaxial halo model.

\begin{figure}
\includegraphics[width=8cm]{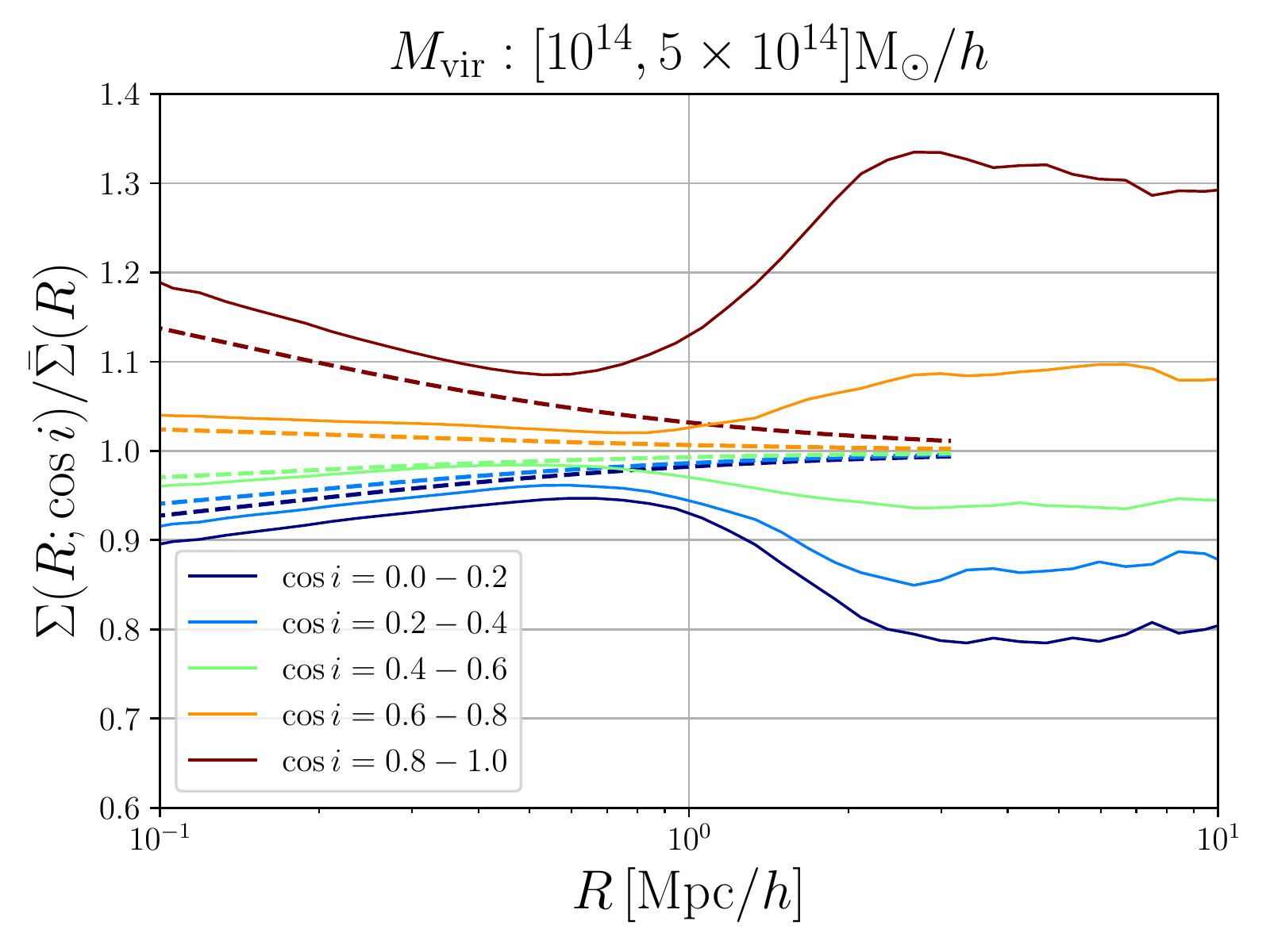}
\includegraphics[width=8cm]{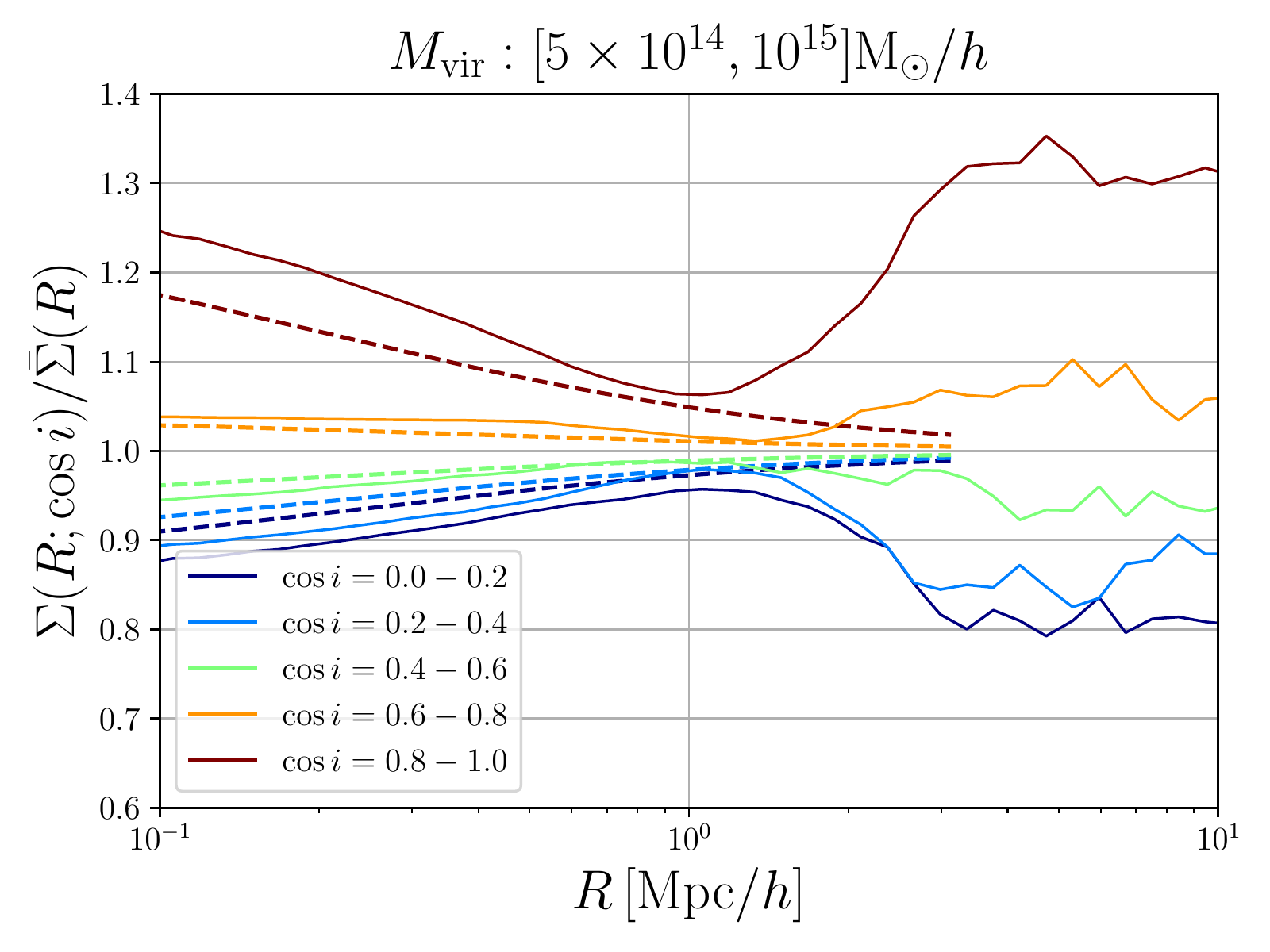}
\includegraphics[width=8cm]{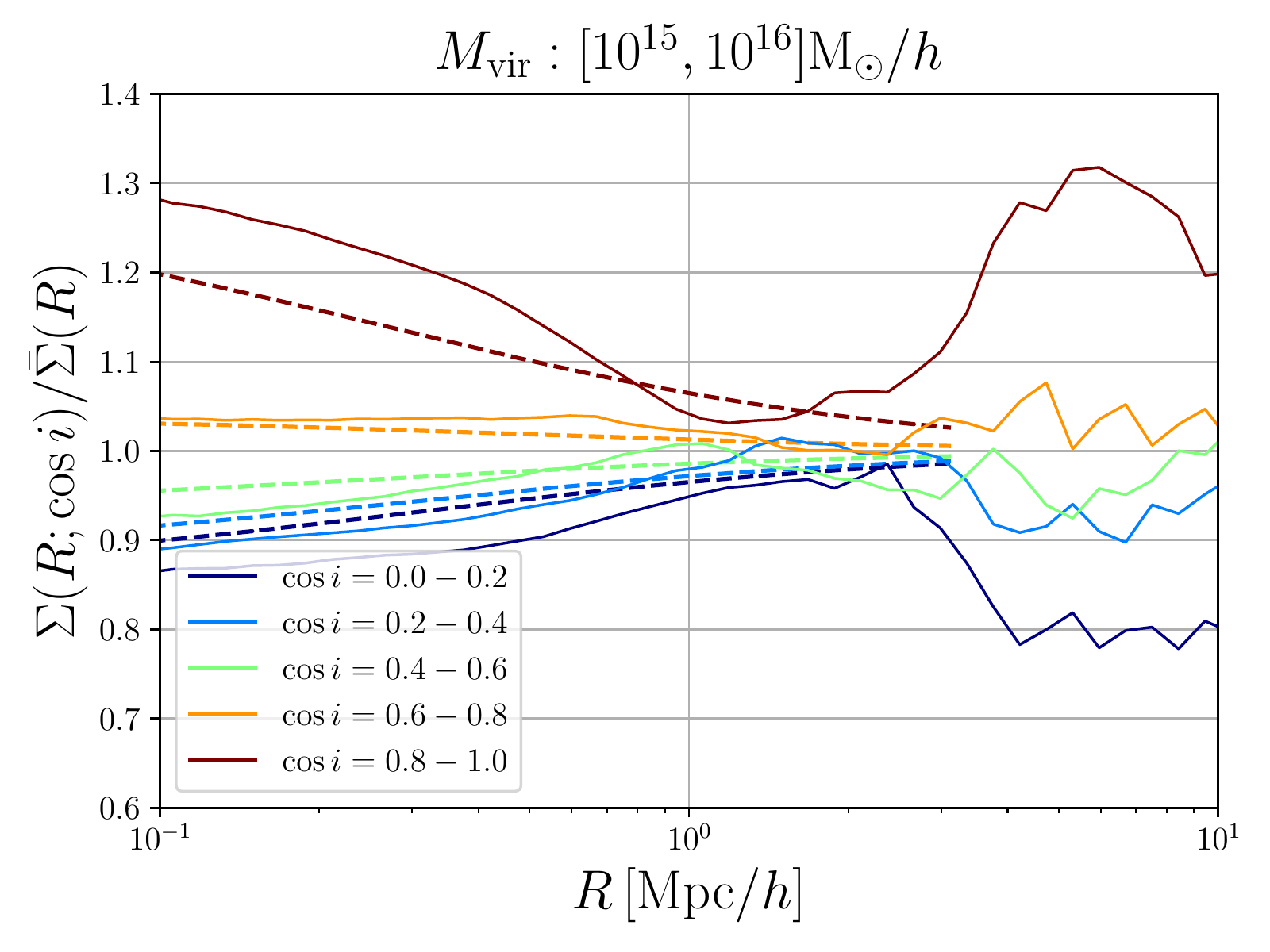}
\caption{The ratios of the surface mass density profiles for different halo
  orientations ({\it solid lines}),
  which are the same as solid lines in
  the lower panels of Figure~\ref{fig:Sigma}
  are compared with those predicted
  by the triaxial halo model of \citet{Jing2002} ({\it dashed lines}).
  The surface mass density profiles of the triaxial halo model are
  computed using the method developed in \citet{Oguri2003} and
  \citet{Oguri2009}.}
\label{fig:trimodel}
\end{figure}

In Figure~\ref{fig:trimodel}, we compare the results from $N$-body
simulations with the triaxial halo model predictions. We find that the
triaxial halo model reproduces the orientation dependence of surface
mass density profiles at small scales ($\la 1\,\Mpc/h$) reasonably well.
While the overall amplitudes of the ratios are slightly
different between the simulations and the triaxial
halo model presumably due to the inaccuracy of the triaxial halo model
especially at high mass end\footnote{\citet{Bonamigo2015} presented
the probability distribution function of axis-ratios with high resolution
$N$-body simulation. Such extension from \citet{Jing2002}
may resolve the discrepancy.},
the model well reproduces the general
trend of the ratios in the one-halo regime, including the radius
dependence of the ratios and the relative behavior of the ratios as a
function of $\cos i$. This comparison supports the idea that the
orientation dependence of surface mass density profiles at small
scales ($\la 1\,\Mpc/h$) can mostly be explained by the halo
triaxiality, and the orientation dependence of surface mass density
profiles at large scales ($\ga 2\,\Mpc/h$) has different origin,
which we will discuss in detail in Section~\ref{sec:origin}.

\subsection{Bias in estimating halo masses}
\begin{figure}
\includegraphics[width=8cm]{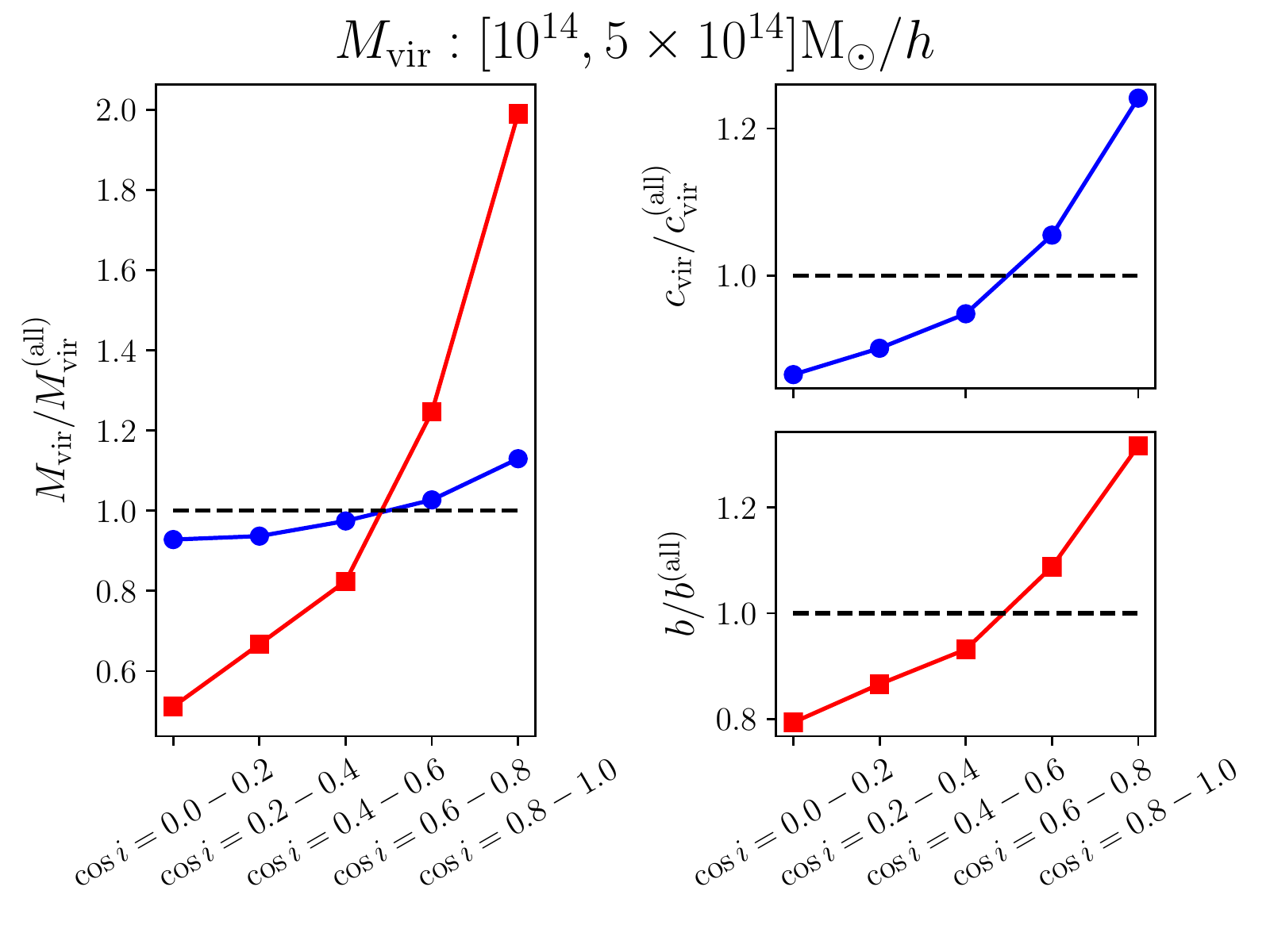}
\includegraphics[width=8cm]{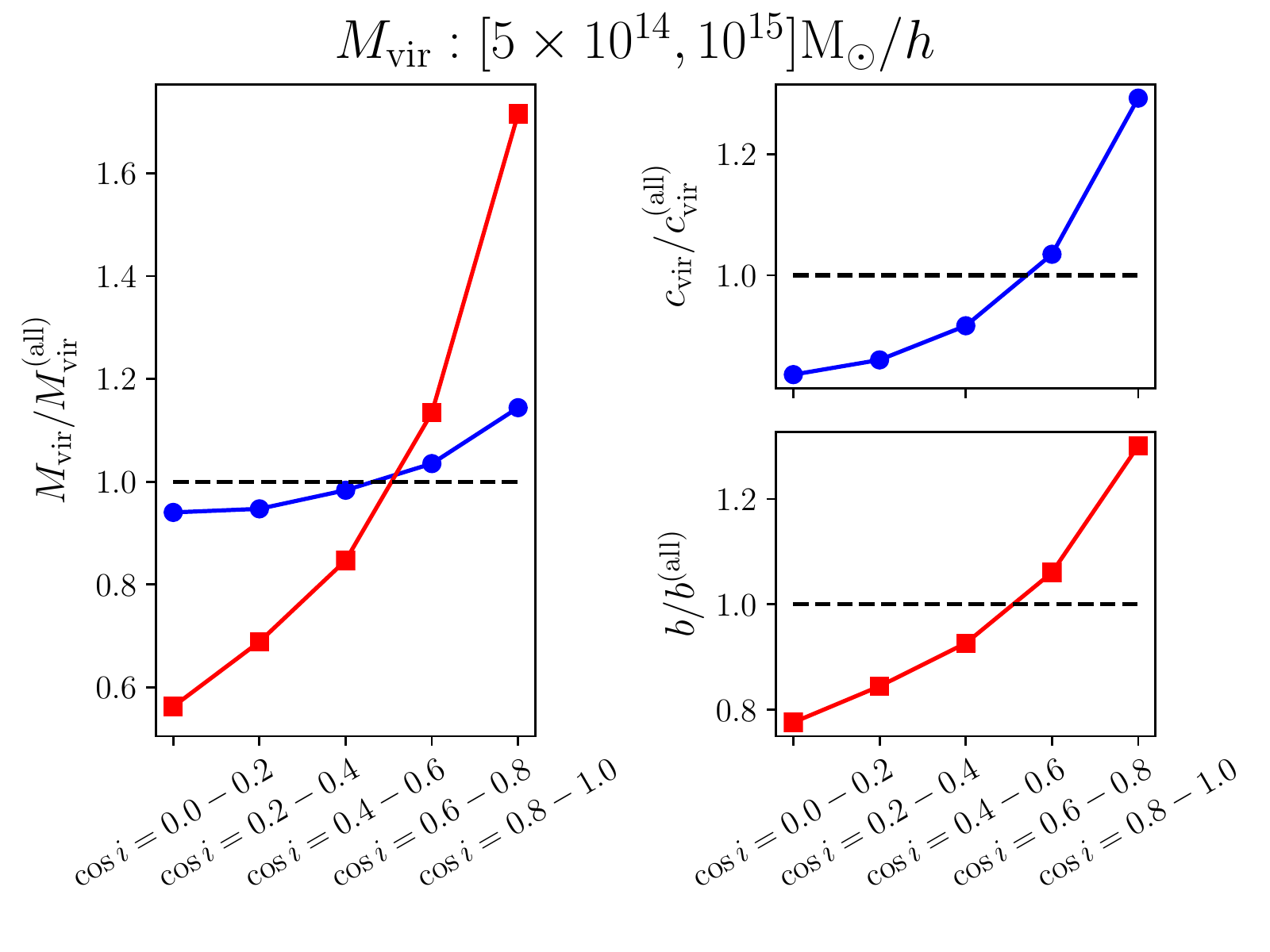}
\includegraphics[width=8cm]{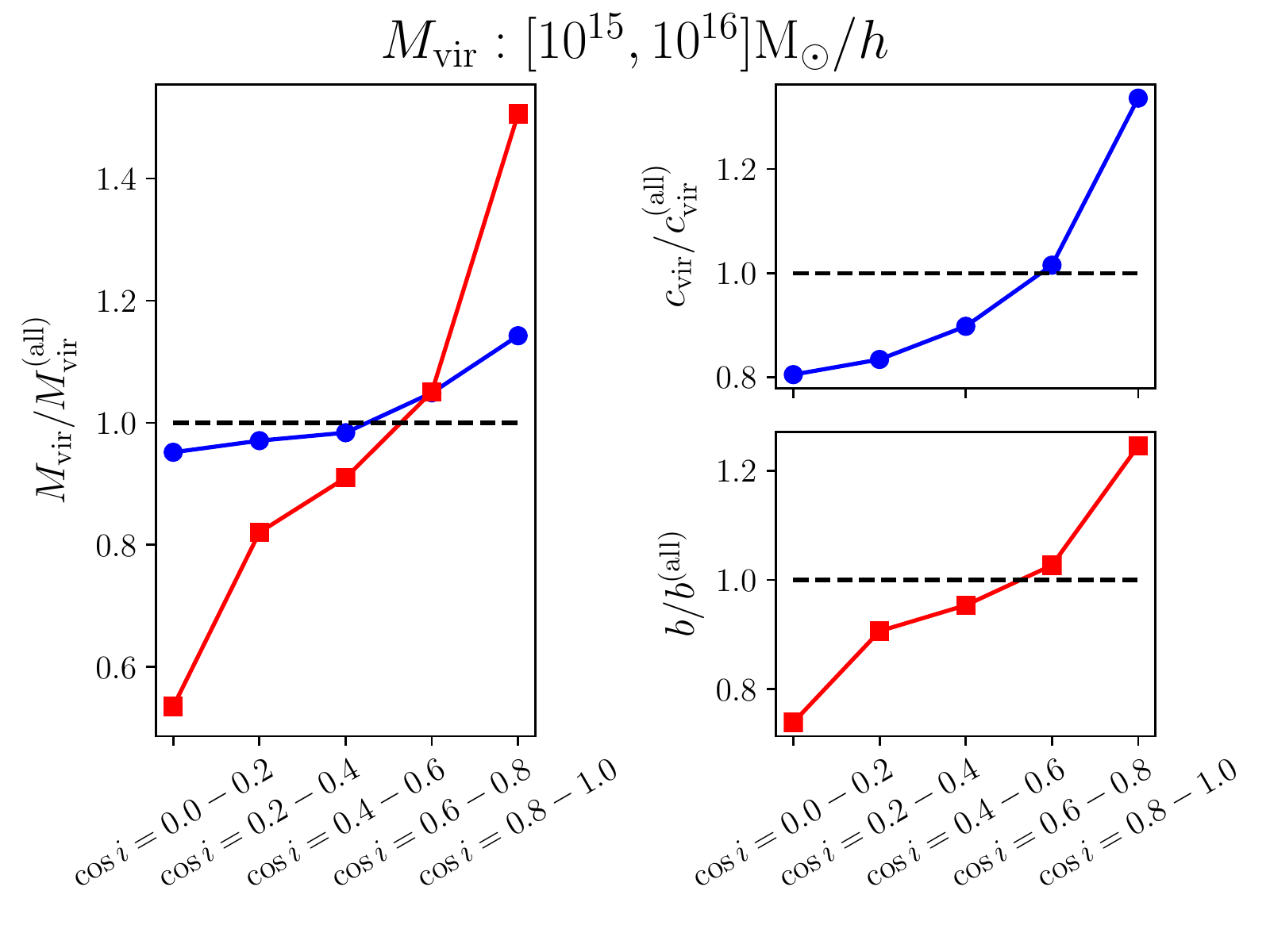}
\caption{For each plot, the best-fit halo mass ({\it left}),
  concentration parameter ({\it upper right}), and halo bias
  ({\it lower right}) are shown as a function of the halo orientation.
  All the values are normalized by the best-fit values for surface mass
  density profiles averaged over all orientations.
  In the left panels, the filled circles show best-fit halo masses
  from the one-halo term, whereas the filled squares show
  the best-fit two-halo mass, i.e., the halo mass inferred from the
  best-fit halo bias using the fitting formula of \citet{Tinker2010}.}
\label{fig:fit}
\end{figure}

The halo orientation changes the amplitude of the surface mass density
profile at all scales. This shift of the amplitude inevitably causes
a bias in cluster properties inferred from an observation of
the surface mass density profile,
if the cluster properties (parameters) such as halo mass
are estimated from the model fitting assuming the spherical symmetry.
We address such a bias by making a hypothetical model fitting of
the surface mass density profiles in simulations with the spherically symmetric
halo model presented in Section~\ref{sec:formalism}.
In doing so, we adopt, as model parameters,
the halo mass $M_\vir$ and the concentration parameter $c_\vir$ for the
one-halo term, and the halo bias $b$ for the two-halo term
for simplicity.
Figure~\ref{fig:fit} shows the estimated parameters for individual orientation bins
relative to the best-fit parameter
values for all haloes denoted as $M^\mathrm{(all)}_\vir$,
$c^\mathrm{(all)}_\vir$, and $b^\mathrm{(all)}$. For the halo mass, in
addition to the halo mass estimated directly from the one-halo term,
we also show the halo mass converted from the estimated halo bias using
the fitting formula in \citet{Tinker2010}, which we call the {\it two-halo mass}.

We find that all the parameters show clear dependence on the halo
orientation. For the parameters of the one-halo term, we find that the
halo mass and concentration parameter can be overestimated or
underestimated up to $\sim 20 \%$ depending on the viewing angle, which
is consistent with the previous analysis based on the triaxial halo
model \citep{Oguri2005}. On the other hand, we find
that the best-fit halo bias from the two-halo term strongly depends on the
halo orientation, which has not been recognized before. We note that
the effect of the halo orientation on surface mass density profiles in
the two-halo regime can well be approximated by a constant shift of
the amplitude over a wide range of radii (see Figure~\ref{fig:Sigma}),
which indicates that the effect can be modeled very well by a change
of the value of the halo bias which is assumed to be scale-independent
in our halo model. We find that the halo bias can be overestimated or
underestimated by up to $\sim 30 \%$ depending on the viewing angle.
This orientation dependence of the halo bias translates into the
orientation dependence of the two-halo mass by up to nearly a factor of
two, which is much larger than that of the halo mass derived directly
from the one-halo term. This large effect of the halo orientation on
the two-halo term highlights the importance of the orientation effect
on the analysis of surface mass density profiles.

The different dependences of the halo mass and two-halo mass on the
halo orientation suggests that it may be possible to break the
degeneracy between the halo mass and orientation based on
detailed measurements of surface mass density profiles in both
the one-halo and two-halo regimes.
We leave the exploration of this possibility for future work.

\section{Origin of the strong orientation dependence at large scales}
\label{sec:origin}

\subsection{Effect of projection thickness}
The intrinsic alignment of dark haloes implies that there is a larger
amount of matter along the major axis of the dark halo, and a smaller
amount of matter along the minor axis for compensation. These two
effects counteract with each other when we consider the projection
along the line-of-sight. This argument also suggests that the
orientation dependence of surface mass density profiles should evolve
considerably as a function of the projection thickness. For instance,
for the halo whose major axis is aligned with the line-of-sight
direction ($\cos i \sim 1$), while they have larger two-halo term
amplitudes than the average when the projection thickness is
sufficiently larger, the two-halo term amplitudes should become
smaller than the average for the smaller projection thickness
because only the matter near the direction of the minor axis,
which is underdense than the average,
is accounted in the projected profile.

We investigate the effect of the projection thickness as follows.
In deriving our results, we project matter over the whole simulation
box, i.e., the projection thickness is $1\,\mathrm{Gpc}/h$.
Here we measure the surface mass density profiles with six different
projection thickness, $\Delta r = 10$, $20$, $50$, $100$, $200$, and
$500\,\Mpc/h$.
First, we divide the whole simulation box into 100
slices with $10\,\Mpc /h$ thickness, and construct a halo number
density field for each slice. Then we compute projected matter density
field for each slice projected over the given thickness listed above.
We can obtain the surface mass density profile by calculating
the cross-correlation between projected halo and matter field with
the fast Fourier transform (FFT). We have 100 pairs of projected halo
and matter density field, leading to measurements of 100 surface mass
density profiles. Finally, we average these 100 measurements.

\begin{figure*}
\includegraphics[width=8cm]{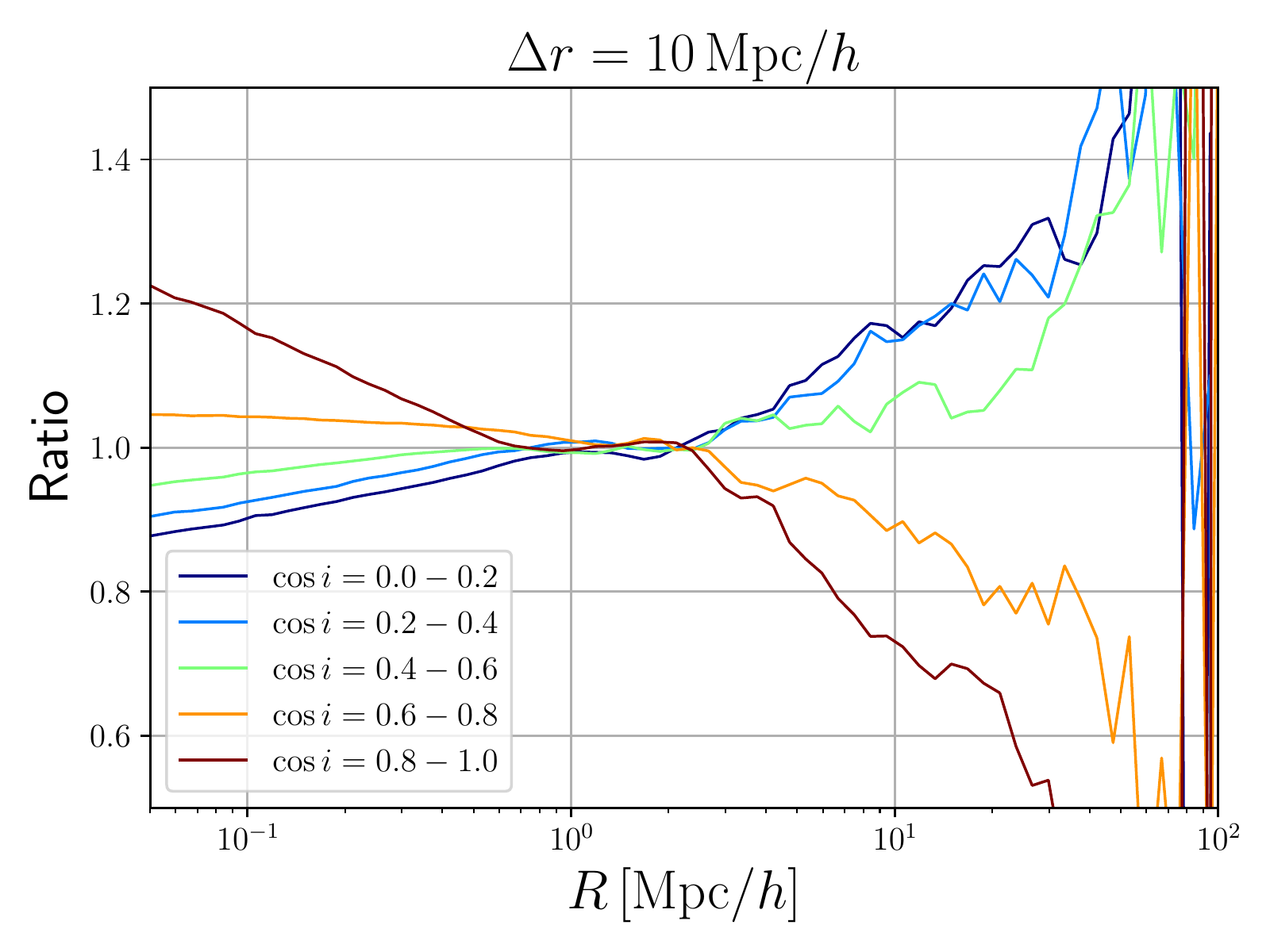}
\includegraphics[width=8cm]{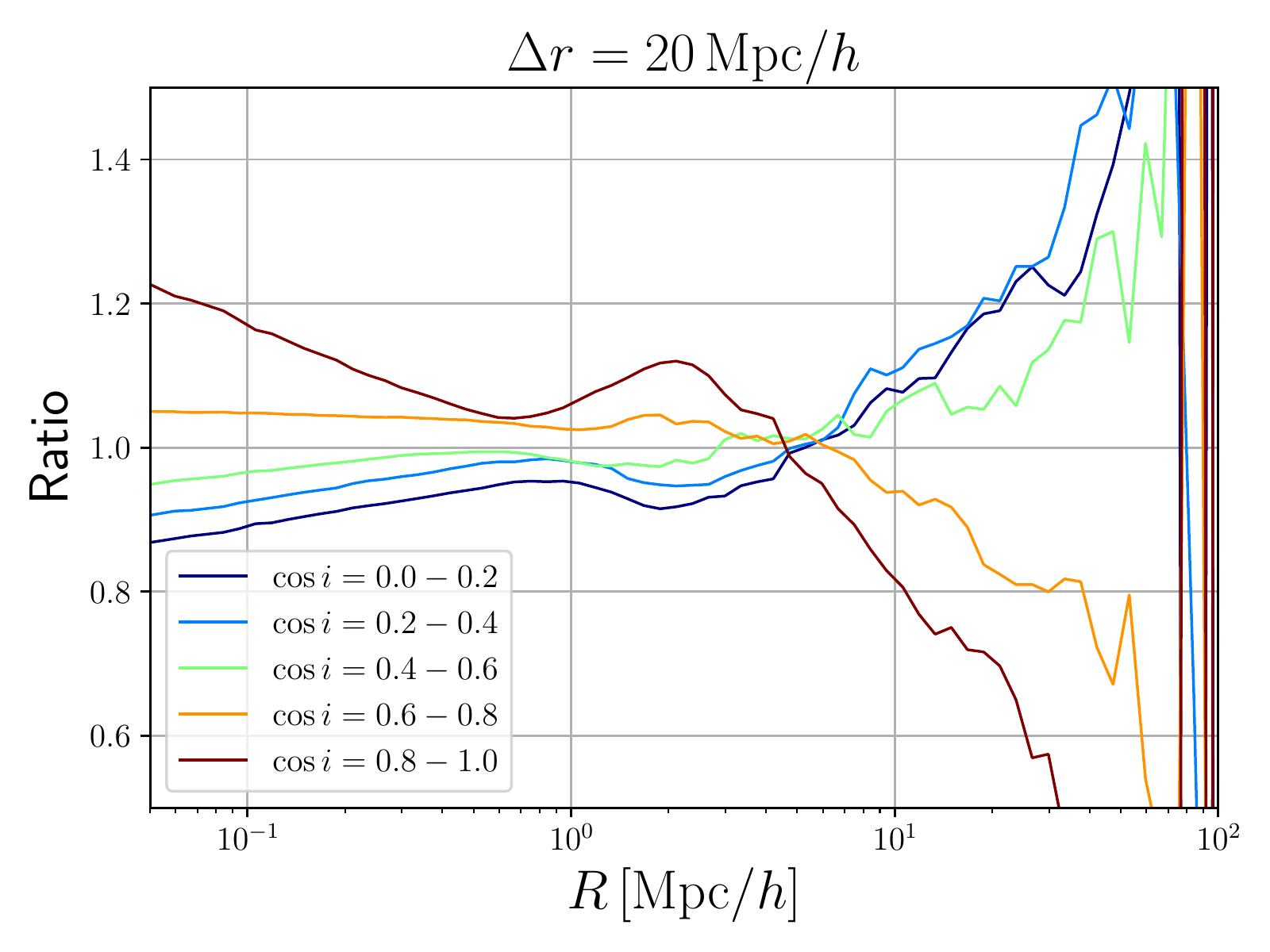}
\includegraphics[width=8cm]{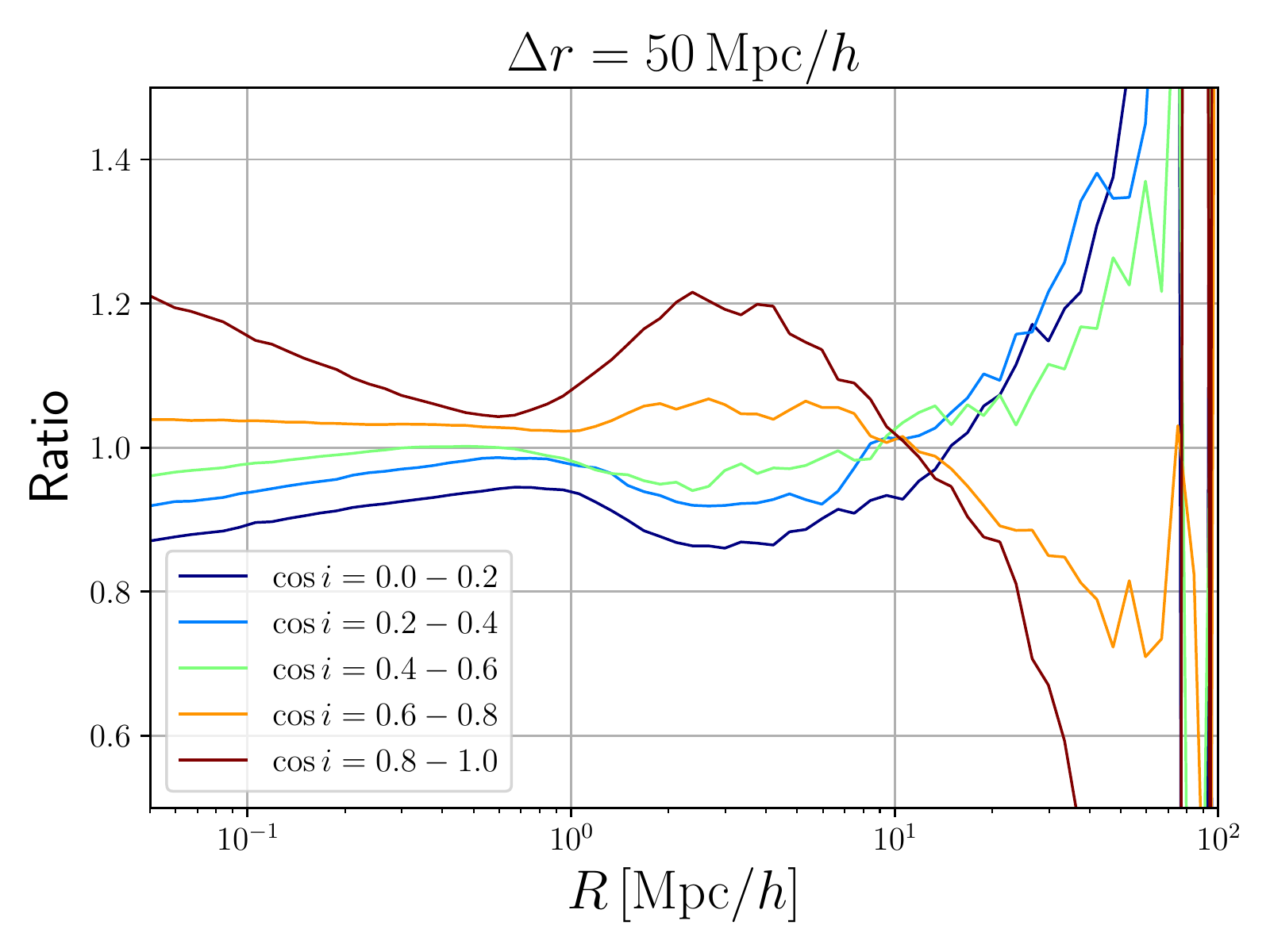}
\includegraphics[width=8cm]{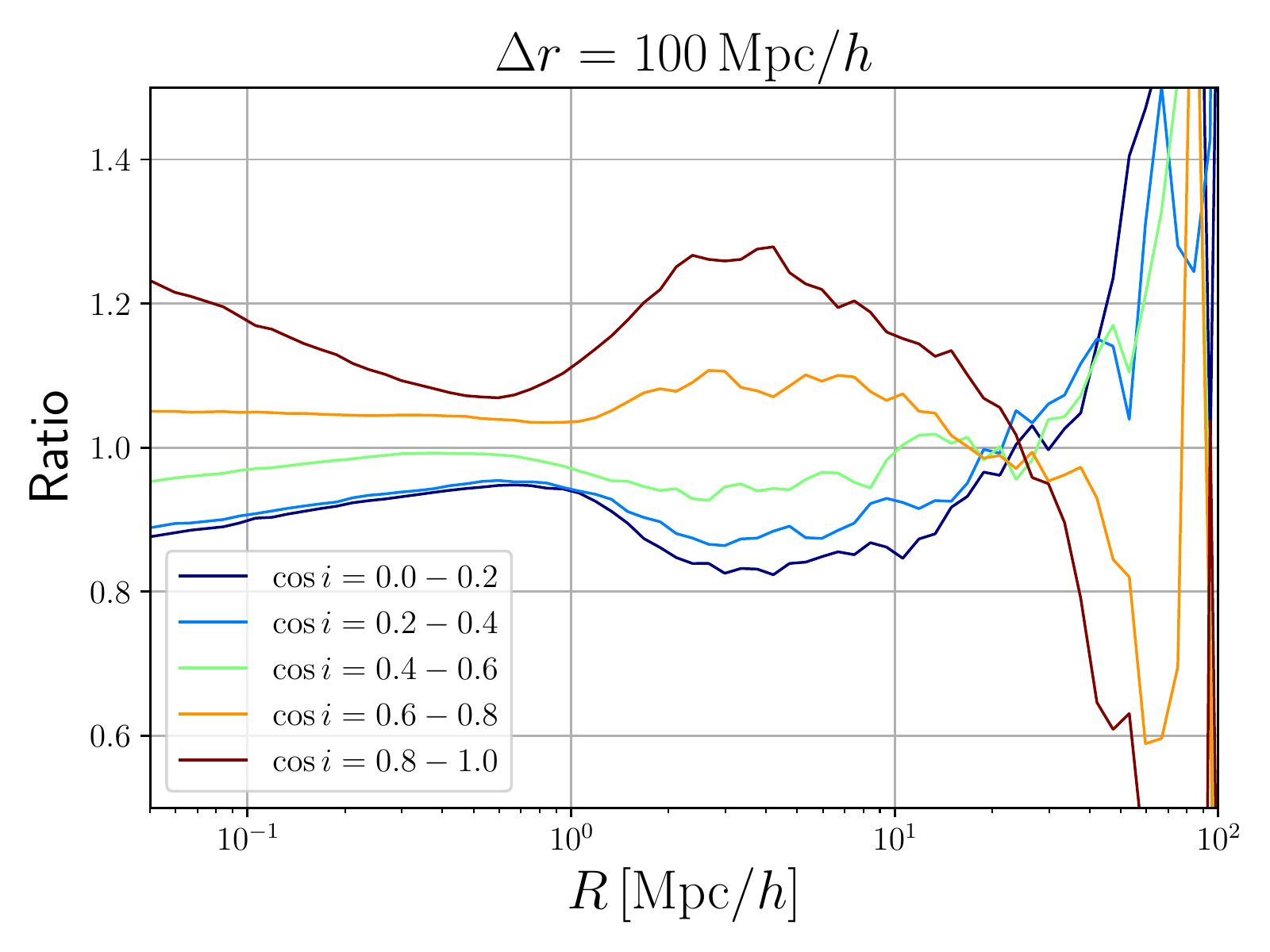}
\includegraphics[width=8cm]{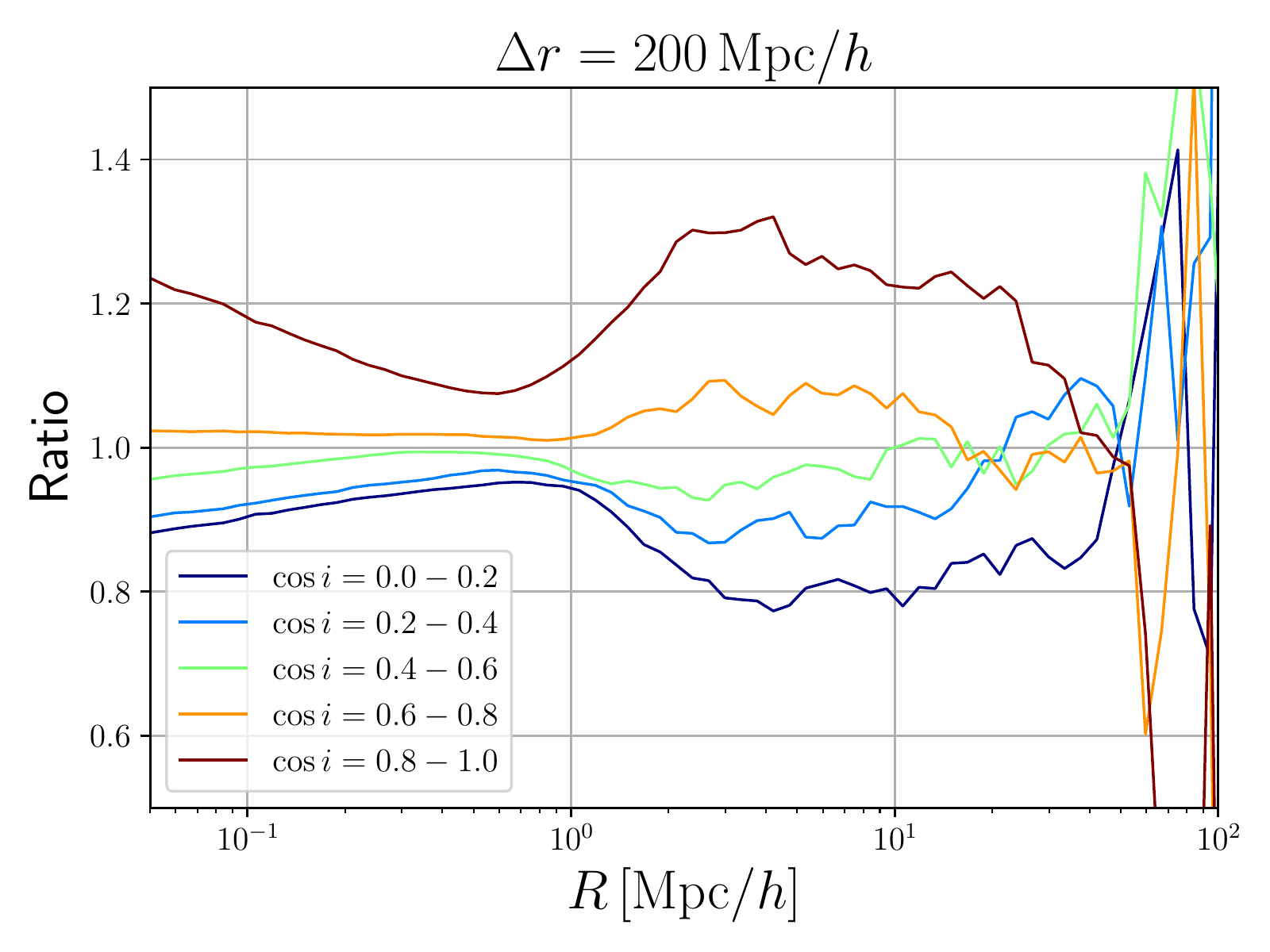}
\includegraphics[width=8cm]{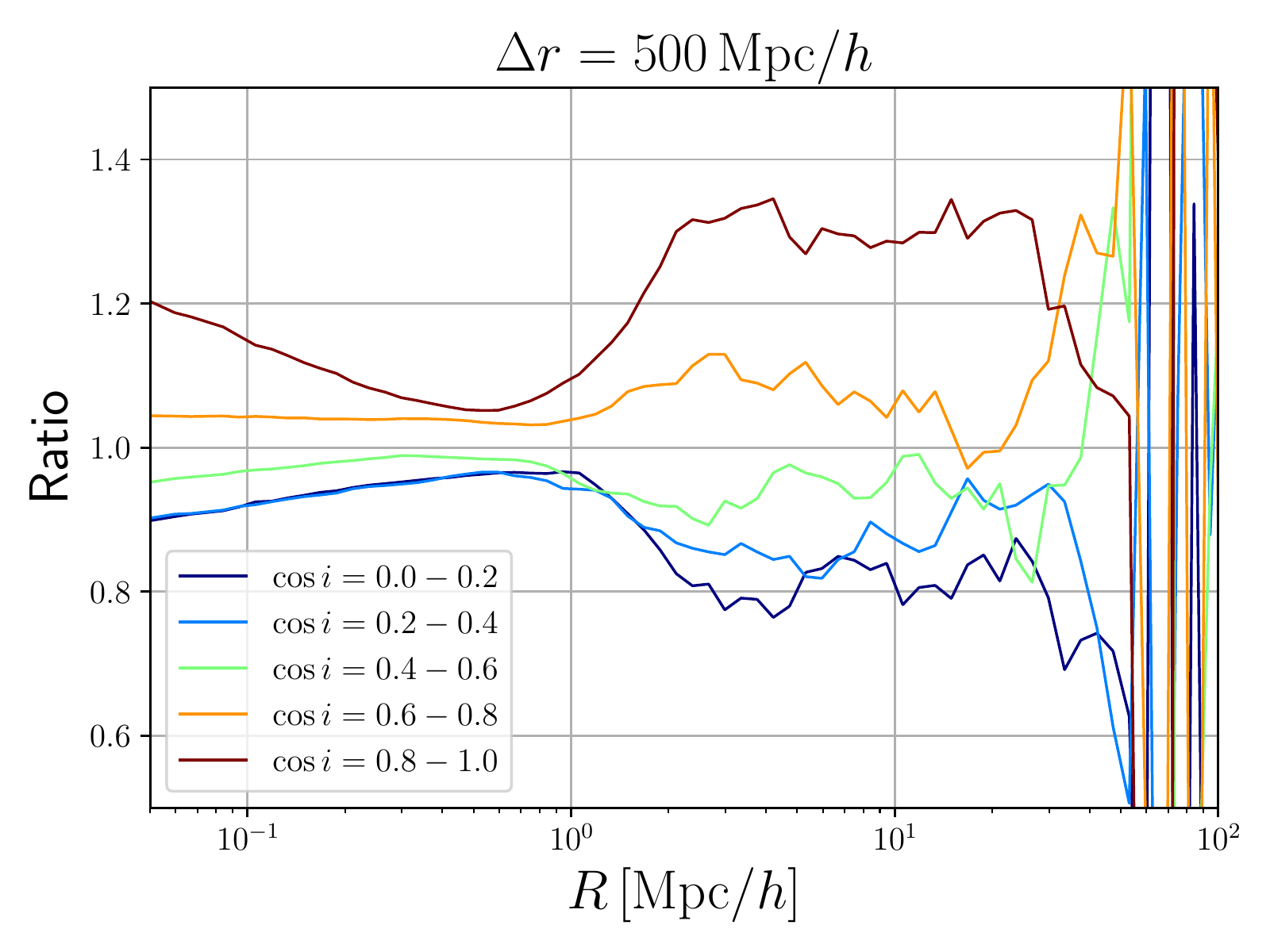}
\caption{Surface mass density profiles for different orientation bins,
  computed using different projection thickness.
  In all panels, the ratios of
  surface mass density profiles for individual orientation bins to
  those for all haloes averaged over orientations are shown.
  In each panel, both the surface mass density profiles for
  individual orientation bins and for all haloes are computed
  with the same projection thickness.
  Here the halo mass range is fixed to $[10^{14}, 5 \times 10^{14}] \Msun /h$.
  We show the results with six different projection thickness,
  $\Delta r = 10$ ({\it upper left}), $20$ ({\it upper right}),
  $50$ ({\it middle left}), $100$ ({\it middle right}),
  $200$ ({\it lower left}), and $500\,\Mpc/h$ ({\it lower right}),
  respectively.}
\label{fig:Sigma_slice}
\end{figure*}

In Figure~\ref{fig:Sigma_slice}, we show surface mass density profiles
with different projection thickness.
As expected, for the case of a thin projection thickness, e.g.,
$\Delta r = 10\,\Mpc /h$, the orientation dependence of surface mass
density profiles at large scales show the opposite trend as compared
with the case of the full projection.
The matter distribution at large separations
is lower than the average, supporting a naive picture
that the matter distribution in the direction perpendicular to the major axis
is indeed in an underdense region such as a void.
\citet{Kawaharada2010} has shown that, from X-ray observations,
the galaxy cluster A1689, whose major axis
is aligned with line-of-sight \citep{Oguri2005,Umetsu2015},
is likely to be surrounded by void regions
in the sky plane (perpendicular to the line-of-sight).
As the projection thickness increases,
surface mass density profiles of aligned haloes ($\cos i \sim 1$)
at large scales gradually become larger because matter
clustered along the major axis starts to be projected.
For the large projection thickness, e.g., $\Delta r = 500\,\Mpc /h$,
the results are quite similar to the one with full projection,
suggesting that the projection thickness of $1\,\mathrm{Gpc}/h$ used
for our main analysis is sufficiently large.
The results in Figure~\ref{fig:Sigma_slice} also imply that
an opening angle of the matter enhancement
along the major axis at large three-dimensional
separations from the haloes is quite wide, because the matter enhancement
remains evident even at such large projected radii up to $100\,\Mpc/h$,
almost the baryon acoustic oscillation scales.

\subsection{Ellipse model}
To gain a better understanding of the results we have shown so far,
we here introduce a phenomenological model to explain the orientation
dependence of surface mass density profiles.
First recall that, without any loss of generality,
the surface mass density profile around
haloes is obtained from a line-of-sight projection of the halo-matter
cross-correlation function, $\xi_\rmhm (r)$ \citep{Mandelbaum2006,Miyatake2015}
\beq
\Sigma(R)=\bar{\rho}_{\mathrm{m0}}\int_{-\infty}^{\infty}\!\!\mathrm{d}z \,
\xi_\rmhm \left( \sqrt{R^2+z^2} \right) ,
\label{eq:sigma_def}
\eeq
where $R \equiv \sqrt{x^2+y^2}$ and the cross-correlation
function is originally given as a function of
the three-dimensional separation $r$,
$r = \sqrt{x^2+y^2+z^2}$, where $z$ is the coordinate in the line-of-sight direction,
$x$ and $y$ are coordinates in the two-dimensional
plane perpendicular to the line-of-sight direction ($z$-direction in our setting).
In the above equation we assumed the axial symmetry around the
light-of-sight direction, i.e., the cross-correlation is obtained from
an angle $\varphi$-average of $(x, y) = R (\cos \varphi, \sin \varphi)$.

When stacking the matter distribution around haloes whose
major axes are aligned with the line-of-sight direction,
it violates the spherical symmetry even in a statistical sense.
That is, the cross-correlation depends on the separation length and its direction:
\beq
\xi_\rmhm (R, z) \ne \xi_\rmhm \left( \sqrt{R^2+z^2} \right) = \xi_\rmhm (r).
\eeq
The persistence of the halo-orientation dependence even at large projection radii would
suggest that the cross-correlation can be approximated by a multipole expansion up to the
quadrupole, in analogous to the redshift-space distortion effect
\beq
\xi_\rmhm (r, \theta) \simeq \xi_\rmhm (r)
\left[ 1 + A \cos 2\theta \right],
\label{eq:xi_quadrupole}
\eeq
where $\cos \theta \equiv z/\sqrt{z^2+R^2}$.
Hereafter, we refer to this model as the quadrupole model.
Our results suggest $A>0$ because we found an enhancement of
the matter distribution along the major axis (i.e. $\theta \simeq 0$).
The assumption that the large-scale matter distribution
around the aligned haloes is described by a sum of the monopole (the first term)
and quadrupole (the second term) might be too simplistic.
However, a possible justification is as follows.
We argue that such a large-scale cross-correlation
between the halo scales, i.e the halo ellipticity at small scales of $\la \Mpc$,
and the large-scale structures up to $100\,\Mpc$ arises from a mode-coupling
between the small- and large-scale Fourier modes in nonlinear structure formation.
As shown in the several work \citep[e.g.,][]{TakadaHu:13,AkitsuTakada:17},
the effect of large-scale modes on small-scale structures
in the deeply nonlinear regime is well modeled by the second-derivatives
of the long-wavelength gravitational potential due to the equivalence principle.
If this is the case, the second-derivative tensor of the scalar potential causes
only up to a quadrupole pattern in the small-scale matter distribution.
We below test this hypothesis.

Motivated by the above picture and also extending
the naive picture, equation~\eqref{eq:xi_quadrupole}, to a more general case,
we assume that the anisotropic cross-correlation function is approximated by
\beq
\xi_\rmhm (R, z) = \bar{\xi}_\rmhm [R_e(R, z)]
\label{eq:xi_e}
\eeq
where $\bar{\xi}_\rmhm (r)$ is the spherically-symmetric,
i.e., averaged over all directions, correlation function and
$R_e$ is chosen to satisfy the following condition:
\beq
\frac{R^2}{q^2} + z^2 = \frac{R_e^2}{q} .
\label{eq:ell}
\eeq
Note that the right hand side of equation~\eqref{eq:ell} is $R_e^2/q$, not $R_e^2$.
This is because we need the {\it mean} radius of the ellipse as the argument
in equation~\eqref{eq:xi_e}. Here, we define the radius $R_e$
such that the area of the circle is equal to
that of the ellipse defined in equation~\eqref{eq:ell}.
In order to satisfy this condition, the additional factor of $q$
is introduced in the right hand side of equation~\eqref{eq:ell}.
The correlation function is given by two variables $(R, z)$,
and we will below treat the ellipticity $q \, (<1)$ as a free parameter
at each separation, i.e., we allow $q$ to vary as a function of
circular radius, $q \left( r= \sqrt{R^2+z^2} \right)$.
This is analogous to the elliptical halo profile
in equation~\eqref{eq:halo_triaxial}.

Once the above model (equation~\ref{eq:xi_e}) is adopted,
the projected surface mass density profile is given as
\beq
\Sigma (R) = \bar{\rho}_\mathrm{m0}
\int_{-\infty}^\infty\!\! \mathrm{d}z \, \xi_\rmhm (R, z; \cos i),
\eeq
where $\cos i$ is the cosine angle between the major axes of haloes and
the line-of-sight direction.
We test how this simplified model can reproduce the results we have shown.

\begin{figure}
\includegraphics[width=8cm]{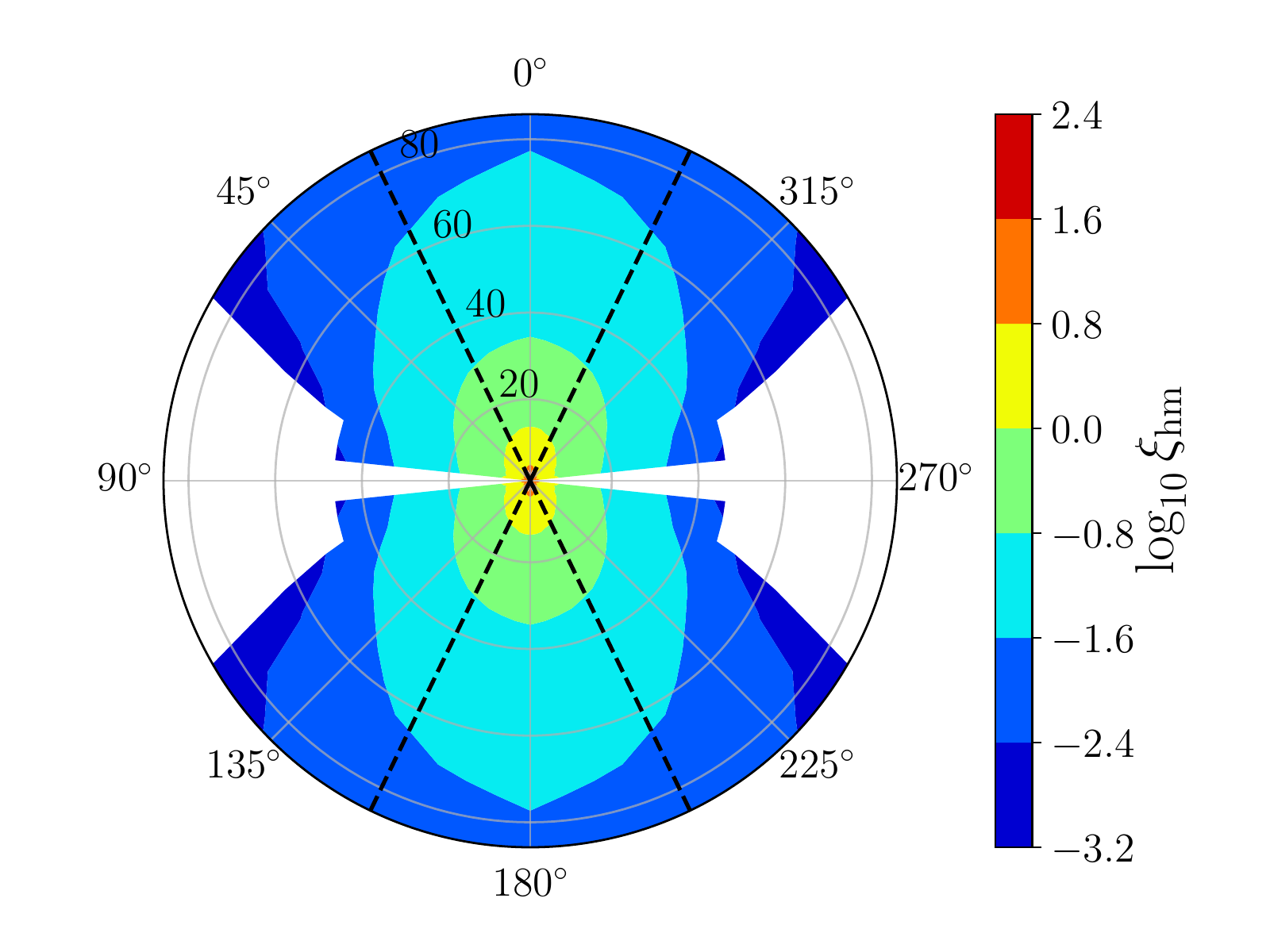}
\caption{The halo-matter cross-correlation $\xi_\rmhm (r, \theta)$
in the two-dimensional polar coordinates.
The radial separation $r$ is taken from the origin and the angular coordinate $\theta$
is defined from line-of-sight. The unit of radius is $\Mpc/h$.
The white regions correspond to the negative values or no data points there.
The black dashed lines show the range of halo major axis direction,
$\Delta i \simeq 26 \degr$ corresponding to $\cos i = 0.9$--$1.0$.}
\label{fig:xi}
\end{figure}

\begin{figure}
\includegraphics[width=8cm]{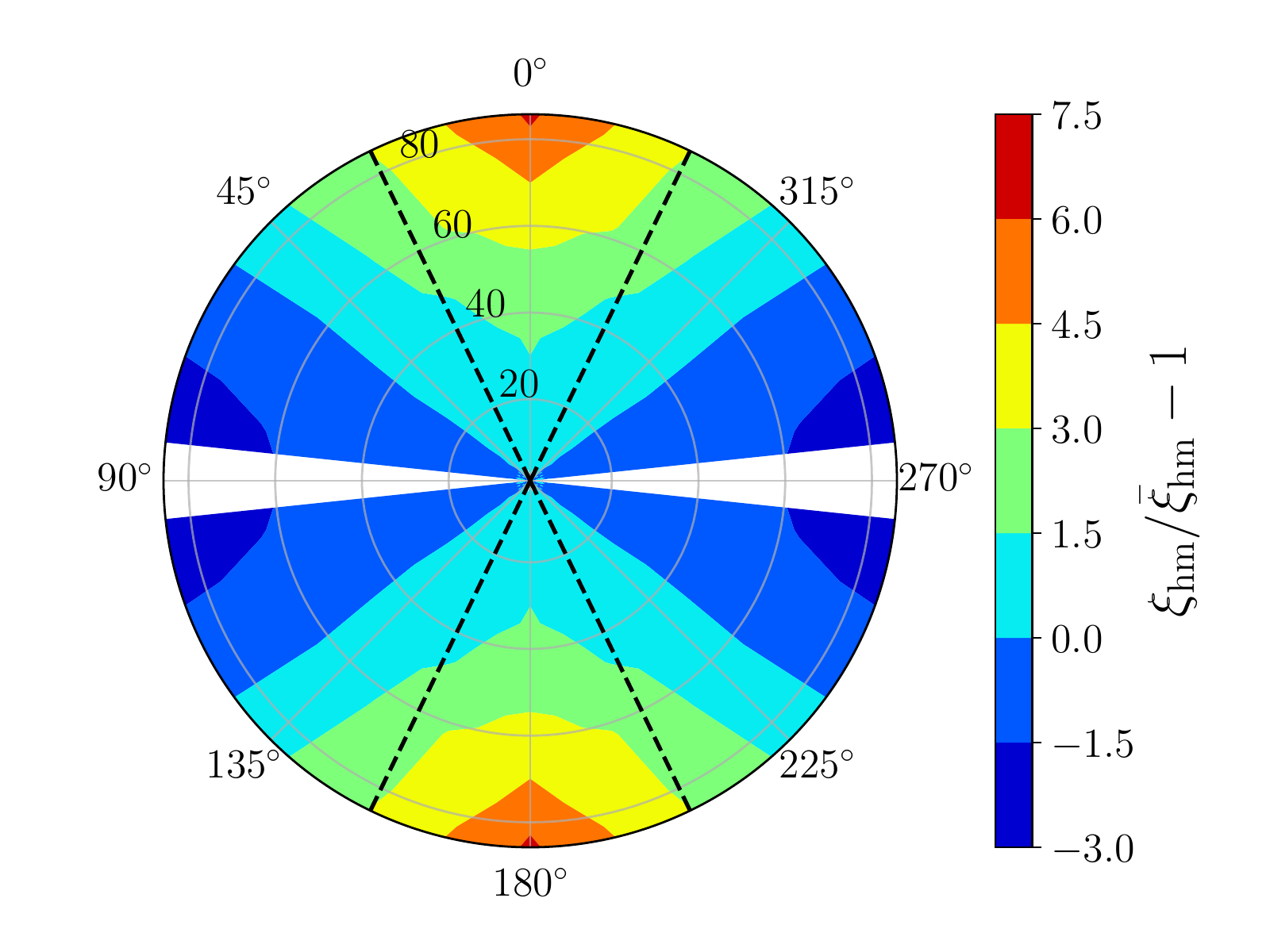}
\caption{The fractional ratio of the halo-matter cross-correlation
in Figure~\ref{fig:xi}, relative to the angle-averaged cross-correlation.
The unit of radius is $\Mpc/h$.
The white regions correspond to no data points there.}
\label{fig:xi_ratio}
\end{figure}

In Figure~\ref{fig:xi},
we show a contour map of the cross-correlation function
in the polar coordinates $(r, \theta)$ converted from $(R, z)$,
measured from the $N$-body simulations.
We selected haloes whose virial masses are in the range of
$[10^{14}, 5 \times 10^{14}] \Msun /h$ and the directional cosine with respect
to line-of-sight satisfies $\cos i > 0.9$.
The line-of-sight ($z$-direction) is taken to the angular direction $\theta = 0 \degr$.
The fractional difference of the cross-correlation
with respect to the angular averaged correlation function, i.e.
$\xi_\rmhm (r, \theta)/\bar{\xi}_\rmhm (r) - 1$, is shown in Figure~\ref{fig:xi_ratio}.
Figure~\ref{fig:xi} clearly shows the anisotropic nature of the
cross-correlation such that the contour is elongated along the major
axis of the halo, up to $80\,\Mpc/h$.
Filaments and/or other haloes connected with filaments
are more likely to exist along the major axis via the tidal field,
which may lead to the elongation.
It appears that the shape of the contour can be
modeled by an ellipse up to the large scales
\citep[see also][]{Schneider2012}. In addition, the fractional difference shown
in Figure~\ref{fig:xi_ratio} also displays a clear quadrupolar feature.
In other words, this figure does not show a clear signature of higher-order
multipole pattern beyond the quadrupole, supporting the argument we gave above.

\begin{figure}
\includegraphics[width=8cm]{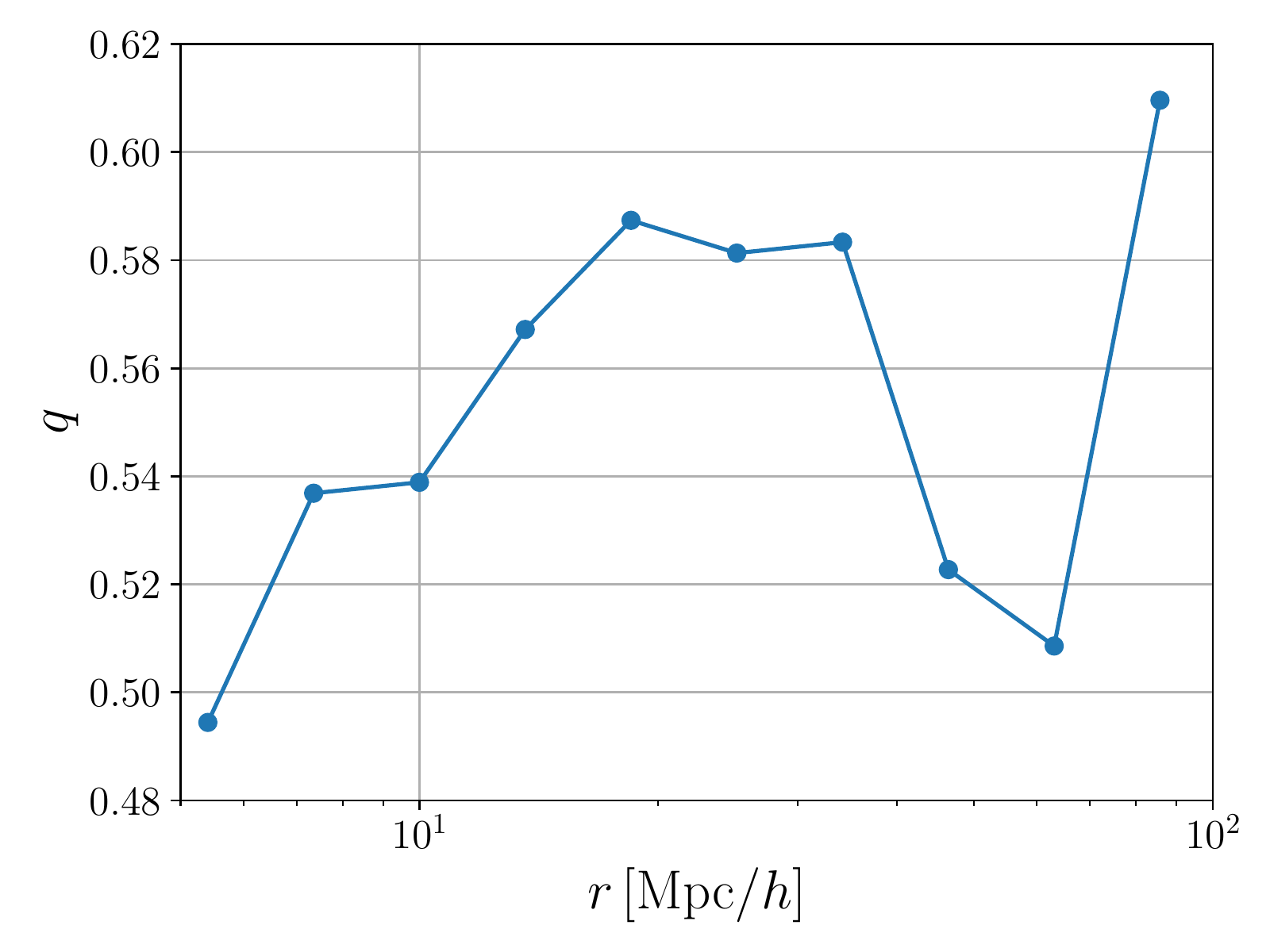}
\caption{The best-fit axis ratio $q$
for the ellipse model prediction (equation~\ref{eq:xi_e}),
where the isoamplitude contour of
the anisotropic cross-correlation $\xi(r, \theta)$ is given by
an ellipse for each radius. We estimate the axis ratio for each of
different separations, denoted by the circle symbols,
based on the method described around equation~\eqref{eq:ell}.}
\label{fig:best_q}
\end{figure}

\begin{figure}
\includegraphics[width=8cm]{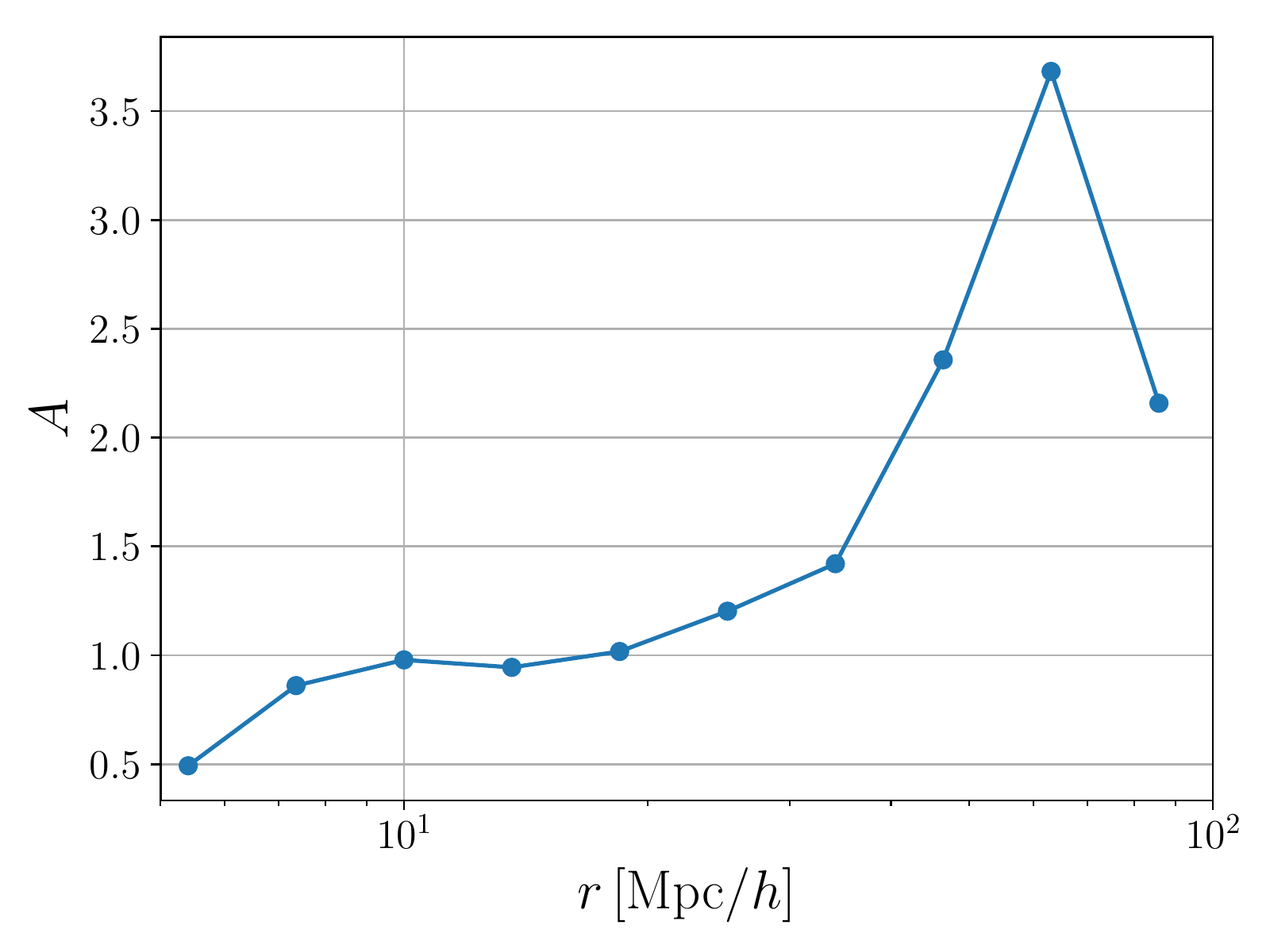}
\caption{The best-fit coefficient $A$
for the quadrupole model prediction (equation~\ref{eq:xi_quadrupole}),
where the anisotropic cross-correlation $\xi(r, \theta)$ can be approximated
as the sum of monopole and quadrupole contributions for each radius.
We estimate the coefficient for each of different separations,
denoted by the circle symbols in a similar manner to the axis ratio $q$.}
\label{fig:best_A}
\end{figure}

\begin{figure}
\includegraphics[width=8cm]{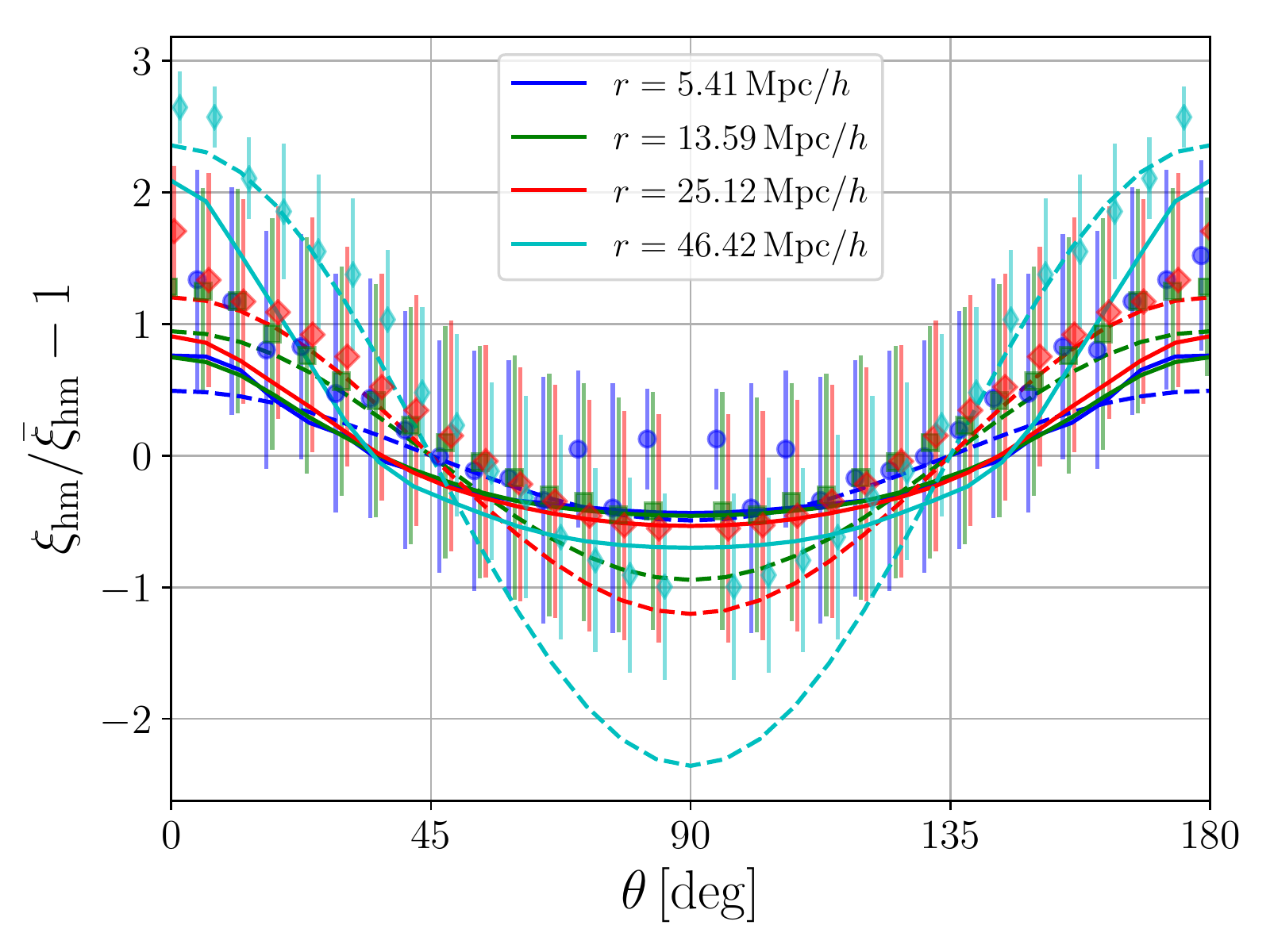}
\caption{The angular dependence of the cross correlation
$\xi_\rmhm (r, \theta)$ is compared with
the ellipse model prediction (equation~\ref{eq:xi_e})
with the best-fit axis ratios in solid lines and the quadrupole model
(equation~\ref{eq:xi_quadrupole}) with the best-fit coefficients
in dashed lines for a given separation.
The results are shown for four different radii,
$r=5.41$ ({\it circles}), $13.59$ ({\it squares}), $25.12$ ({\it square diamonds}),
and $46.42\,\Mpc/h$ ({\it thin diamonds}).
The error bars correspond to the standard deviation over the 24 realizations.}
\label{fig:ell_model}
\end{figure}

In order to quantify this behavior,
we fit the isoamplitude contour of $\xi_\rmhm (r, \theta)$
by the ellipse model equation~\eqref{eq:xi_e}
for each polar radius with varying the parameter $q$.
In Figure~\ref{fig:best_q}, the best-fit axis ratio is
shown as a function of the radius, for radial bin $r > 5\,\Mpc/h$.
The best-fit axis ratios slightly depend on the
radius, but for a wide radius range the best-fit axis ratio is $q \sim 0.55$.
For radial bins smaller than $5\,\Mpc/h$,
the least chi-square method does not converge.
Similarly, we estimate the best-fit coefficient $A$
in the quadrupole model (equation~\ref{eq:xi_e}).
Figure~\ref{fig:best_A} shows the best-fit coefficient $A$
for radial bin $r > 5\,\Mpc/h$.
The coefficient increases with radius,
which indicates the variation with respect to the angle
is strengthened for outer regions.
In Figure~\ref{fig:ell_model}, we show the comparison of
the best-fit ellipse or quadrupole models with
the measured cross-correlation functions at different radii.
We find that the both models can fairly well reproduce the angular dependence
except the case of the largest radial bin,
$r = 46.42 \,\Mpc/h$. For the ellipse model,
that inconsistency may be caused by the
inaccurate averaged cross-correlation measurements in simulations due
to the limited size of the simulation box.
For the quadrupole model,
the slightly larger deviation implies that the model cannot describe the higher-order multiple
contributions such as hexadecapole dependence that exist in the simulation results.
In any case, we expect that this ellipse model is useful for analyzing
the orientation dependence of surface mass density profiles.

\section{Discussions}
\label{sec:discussions}
In this Section, we discuss potential impacts of our results on the
analysis of observational data. If the halo sample is selected without
any orientation bias, i.e., if the distribution of major axes is
completely random with respect to the line-of-sight direction, the
orientation dependence of surface density mass profiles is averaged out
so that we can simply ignore the orientation dependence derived in
this paper. For example, it is natural to expect that a sample of
normal galaxies is unbiased with respect to their orientation with
respect to the line-of-sight direction\footnote{The orientation bias
might be relevant even for a galaxy sample, e.g., if galaxies are
selected based on the aperture photometry with the aperture size
smaller than galaxy sizes. Our results presented in this paper
indicates that it is important to explicitly check the possible
orientation bias for any sample used for the analysis of surface
mass density profiles.}.
However, in some cases sample selections
inevitably introduce the orientation bias, and our results indicate
that correlation analysis of these samples can be significantly
affected by the strong orientation dependence of surface mass density
profiles. Here we discuss several cases where the orientation bias
can be significant.

\subsection{Lensing selected clusters and galaxies}
It has been known that cluster samples selected based on gravitational
lensing exhibits the strong orientation bias. \citet{Hennawi2007}
showed that clusters that exhibit strong lensing features tend to have
their major axes aligned with the the line-of-sight direction, with
the median angle of $\cos i = 0.67$. \citet{Oguri2009} argued that the
orientation bias is even stronger for strong lensing clusters with
large Einstein radii \citep[see also][]{Meneghetti2010}.
\citet{Hamana2012} showed that the similar orientation bias exists
also for weak lensing selected clusters, i.e., clusters selected from
peaks in weak lensing mass maps. Our results predict that surface mass
density profiles of these strong and weak lensing selected clusters
have larger amplitudes at large radii as compared with those for other
cluster samples. We note that the similar orientation bias is also
known to exist for galaxy-scale strong lensing \citep{Rozo2007}.

\subsection{Optically selected clusters}
Recent wide-field photometric surveys with multiple passbands allow us
to identify clusters of galaxies efficiently based on clustering of
red-sequence galaxies. However, due to the projection of red-sequence
galaxies along the line-of-sight, this method may preferentially
select clusters whose major axes are aligned with the line-of-sight
direction. \citet{Dietrich2014} explicitly studied the orientation
bias of optically selected clusters using mock galaxy catalogs, and
argued the existence of the orientation bias particularly for
relatively less massive clusters.

Given the large number of optically selected clusters, high
signal-to-noise measurements of surface mass density profiles with
stacked weak lensing are possible, particularly in ongoing and future
optical imaging surveys. Thus it is of great importance to accurately
quantify the orientation bias of optically selected clusters in order
to take proper account of the orientation dependence of the surface
mass density profile for the accurate mass calibration.

\begin{figure}
\includegraphics[width=8cm]{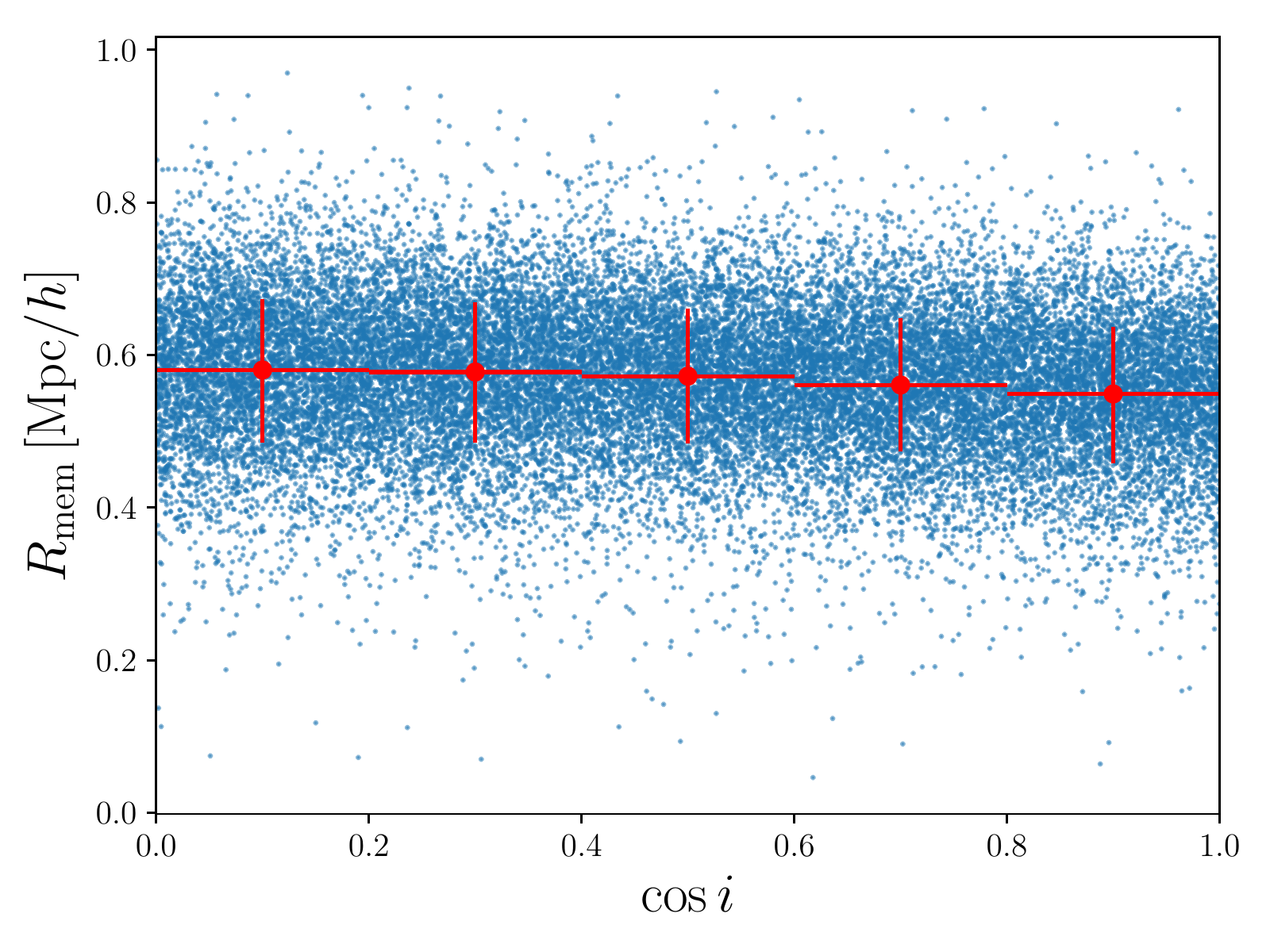}
\caption{The correlation between the average projected separation of
member galaxies $R_\mathrm{mem}$ \citep[see][]{Miyatake2016} and the
orientation of the halo with respect to the line-of-sight
direction. Small filled circles show the full distribution, whereas
large filled circles with error bars show median values and the $16\%$
and $84\%$ percentiles for individual orientation bins.}
\label{fig:Rmem}
\end{figure}

\citet{Miyatake2016} claimed that clusters with member galaxies
located closer to the center on average have large halo bias, which
was ascribed to the detection of the {\it assembly bias}.
However, if the average member galaxy separation correlates with the
halo orientation, such a correlation can mimic the assembly bias
signal and can explain the behavior observed by \citet{Miyatake2016}.
We check this possible correlation using our simulations. We focus
only on haloes with masses in the range of $[10^{14}, 5 \times 10^{14}] \Msun /h$.
We derive the distribution of member galaxies in
simulations using subhaloes with virial masses larger than $5 \times 10^{11} \Msun/h$.
For each dark halo, we use member galaxies within
$1\,\Mpc/h$ from the halo center in the two dimensional space
perpendicular to the line-of-sight direction, and $r_\parallel < 50\,\Mpc/h$
along the line-of-sight direction.
The former and latter ranges correspond to the projected cluster
radius and the photometric redshift error, respectively.
Thus, we compute the average distance of member galaxies,
$R_\mathrm{mem}$, in the same manner as in \citet{Miyatake2016}.
Figure~\ref{fig:Rmem} indicates that the average member galaxy
separation depends very little on the halo orientation, which is
insufficient to explain the large variation of the halo bias found in
\citet{Miyatake2016}.
Hence, if the cluster sample used in \citet{Miyatake2016}
suffers from the orientation bias,
it would be likely from details in the cluster finding algorithm, rather than
the simplified method using a catalog of subhaloes in each halo region.
\citet{Busch2017} addressed how the assembly bias signal
can be reproduced due to the projection effect
using simulated mock catalogs.
They claimed that the assembly bias signal appears when red-sequence galaxies
within $250 \, \Mpc /h$ from the cluster center in the line-of-sight direction
are misidentified as member galaxies.

\subsection{Quasars}
Clustering of quasars has been used to infer their host halo
masses. However, given the small number of quasars, projected
statistics have been used in most cases. Our results indicate that the
orientation effect can be important for the clustering analysis of
quasars if the jet axis of quasar is aligned with the halo
orientation. For instance, \citet{Zhang2009} used $N$-body simulations
to show that the spin and {\it minor} axis of haloes are well
aligned. Thus surface mass density profiles around quasars may also
exhibit strong orientation dependence, if the spin axis of haloes is
aligned with the jet axis.

If the argument above holds, it implies that type 2 (obscured) quasars
have higher apparent clustering signals than type 1 (unobscured)
quasars, even if their underlying host halo masses are similar. In fact,
recent studies, using projected correlation functions as well as
cosmic microwave background lensing measurements, have shown that
type 2 quasars are more strongly clustered than type 1 quasars
\citep[e.g.,][]{Hickox2011,Donoso2014,DiPompeo2016}. These
observations have been used as a counterargument to the so-called
unified model in which the difference between type 1 and type 2
quasars are explained solely by their viewing angles. Our results may
imply that this issue needs to be revisited taking account of the
possible orientation bias of quasars.

\section{Conclusion}
\label{sec:conclusion}
In this paper we have studied how the surface mass density profiles
of haloes vary with the orientation of halo major axis
with respect to the projection direction, using a suite of $N$-body simulations.
Dark haloes are known to be highly non-spherical by nature, and this non-sphericity
leads to variations in the surface mass density profiles
depending on the projection direction with respect to the halo shapes.
We have found that, if projecting the matter distribution around haloes
whose major axes are aligned with the line-of-sight (projection) direction,
the average surface mass density profiles have larger
amplitudes over all scales we studied,
from $0.1$ to $100\,\Mpc/h$, than surface mass density profiles
averaged over all orientations.
The orientation bias at small projected separation
in the one-halo regime is easily understood by the triaxiality
of dark matter halo profile. However, we have shown that the orientation bias
at large separations up to $100\,\Mpc/h$ remains significant.
This is opposite to what one would naively expect;
since the mass distribution in the direction perpendicular
to the major axis (i.e. along the minor or intermediate axis) tends to be
in an underdense region such as void,
we naively expect that the surface mass density projected at
such a large separation is mainly from the underdense region,
and therefore has smaller, instead of larger, amplitude than the average.
We showed that the mass distribution
in the direction of the major axis still affects the projected mass
density profile even at large projection separations,
and the opening angle of the enhanced mass distribution,
viewed from the halo center, is quite wide.

Using the halo model, we have quantified how the halo-orientation dependence of
surface mass density profiles causes a bias in parameters such as halo mass,
concentration parameter and halo bias if those are estimated
from the surface mass density profiles as in the weak lensing observables.
We have found that the halo bias can be overestimated or
underestimated by up to $\sim 30\%$ depending on the viewing angle
with respect to the major axis. The orientation dependence of
the halo bias from the two-halo term translates into the difference of
the two-halo mass by up to a factor of two, which is much larger than
the orientation dependence of the inferred halo mass from the one-halo
term, which is $\sim 20\%$ at most.

In order to understand the origin of the orientation dependence at
large scales, we have studied the average matter distribution around haloes,
i.e. the halo-matter cross-correlation function,
computed with selecting haloes whose major axes are
aligned with line-of-sight, because the surface mass density profile is obtained
from a projection of the halo-matter cross-correlation function at a fixed projected radius.
The cross-correlation displays a clear quadrupole pattern in the mass distribution up to
$\sim 100\,\Mpc/h$, with a significant opening angle of the mass enhancement
along the major axis direction.
The isoamplitude contour of the halo-matter cross-correlation
at large scales is approximated by an ellipse with a axis ratio of $q \sim 0.55$.
The strong cross-correlation up to large scales
implies that the large-scale, anisotropic matter distribution,
which is referred to as the large-scale {\it tidal} field, is the origin of
halo shapes at small scales due to the mode coupling in nonlinear structure formation.
It would be interesting to study whether the cross-correlation exists
out to even larger separations such as baryon acoustic oscillation scales
and remains or disappears, e.g. due to the mass conservation at very large scales.
This is our future study, which will be presented elsewhere.

Our results have various implications.
If a sample of large-scale structure tracers such as galaxies, clusters and quasars
is affected by selection effects depending on the orientation of host haloes,
the analysis of the sample assuming spherical symmetry
can lead to a significant bias in estimating their halo masses
or halo biases. We have discussed lensing and optically selected
clusters as well as quasars for possible examples of samples with the
significant orientation bias.

There are other observables that might be affected by the halo orientation bias.
First, if a selection of spectroscopic galaxies is somehow affected
by halo shapes or filaments in large-scale structure,
the redshift-space clustering of galaxies
would be modified by the halo orientation bias. It is recently shown that
the large-scale anisotropic matter distribution, i.e. the tidal field,
causes an anisotropic clustering pattern in the redshift-space
galaxy distribution due to the nonlinear mode coupling
\citep{AkitsuTakadaLi:17,AkitsuTakada:17}
in addition to the redshift-space distortion and
the cosmological distortion such as the Alcock-Paczynski effect.
This could cause a bias in cosmological parameters estimated from
the measured redshift-space clustering, if the orientation bias is ignored.
Our results also imply that the peculiar velocity field in large-scale structure,
which causes the redshift-space distortion, would have a strong correlation
with shapes of haloes \citep[e.g.][]{Okumuraetal:17}.
Another effect is the intrinsic alignment,
which is one of the major systematic effects in weak lensing measurements.
Our results show that the large-scale matter distribution
has a strong correlation with shapes of haloes.
If a sample of galaxies used in the shape measurements
for weak lensing has a correlation with shapes of haloes,
the sample would cause an intrinsic alignment contamination
to the weak lensing measurement. These would be interesting to study,
and our methods presented in this paper will be useful for such a study.

\section*{Acknowledgements}
We thank H. Miyatake, M. Shirasaki, R. Takahashi, and T. Oogi
for useful discussions and K. Umestu for comments
on an earlier version of the manuscript.
KO is supported by Research Fellowships of the Japan Society for the
Promotion of Science (JSPS) for Young Scientists and Advanced Leading
Graduate Course for Photon Science.
TO acknowledges support from the Ministry of Science and Technology of Taiwan
under the grant MOST 106-2119-M-001-031-MY3.
TN acknowledges financial support
from Japan Science and Technology Agency (JST) CREST Grant Number JPMJCR1414.
This work was supported by JSPS Grant-in-Aid for JSPS Research Fellow
Grant Number JP16J01512.
This work was supported in part by World Premier International
Research Center Initiative (WPI Initiative), MEXT, Japan, and JSPS
KAKENHI Grant Number JP26800093, JP23340061, JP26610058, JP15H03654,
JP15H05887, JP15H05892, JP15H05893, JP15K21733, and 17K14273.
Numerical simulations were carried out on Cray XC30
at the Center for Computational Astrophysics,
National Astronomical Observatory of Japan.



\bibliographystyle{mnras}
\bibliography{main}



\appendix


\bsp 
\label{lastpage}
\end{document}